\documentclass[fleqn,usenatbib]{mnras}

\usepackage{latexsym,mathrsfs,amssymb,bm}
\usepackage[tbtags]{amsmath}
\usepackage[T1]{fontenc}
\usepackage{ae,aecompl,times}
\usepackage{epsfig}
\usepackage[dvipsnames]{xcolor}
\usepackage{multirow}
\usepackage{tikz,array,color,float}
\usepackage[utf8]{inputenc}
\usepackage[T1]{fontenc}
\usepackage{graphicx}
\usepackage{caption}
\usepackage{multirow}
\newcommand{\Mpch}{\,h^{-1}{\rm Mpc}}

\newcommand{\kpch}{\,h^{-1}{\rm kpc}}

\title[Convergence maps with the MTNG simulations]{The MillenniumTNG Project: The impact of baryons and massive neutrinos on high-resolution weak gravitational lensing convergence maps}

\author[F. Ferlito et al.]{%
\parbox{0.98\textwidth}{
Fulvio Ferlito$^{1}$\thanks{E-mail: \href{mailto:ferlito@mpa-garching.mpg.de}{ferlito@mpa-garching.mpg.de}},
Volker Springel$^{1}$,
Christopher T. Davies$^{2}$,
C\'esar Hern\'andez-Aguayo$^{1,3}$, \\
R\"udiger Pakmor$^{1}$,
Monica Barrera$^{1}$,
Simon D. M. White$^{1}$, 
Ana Maria Delgado$^{4}$, \\
Boryana Hadzhiyska$^{4,5,6}$,
Lars Hernquist$^{4}$,
Rahul Kannan$^{7,4}$,
Sownak Bose$^{8}$
and Carlos Frenk$^{8}$
}
\\%
\\%
$^{1}$Max-Planck-Institut f\"ur Astrophysik, Karl-Schwarzschild-Str. 1, D-85748, Garching, Germany\\%
$^{2}$Faculty of Physics, Ludwig-Maximilians-Universit\"at, Scheinerstr. 1, 81679 Munich, Germany\\%
$^{3}$Excellence Cluster ORIGINS, Boltzmannstrasse 2, D-85748 Garching, Germany\\%
$^{4}$Center for Astrophysics | Harvard $\&$ Smithsonian, 60 Garden St, Cambridge, MA 02138, USA\\%
$^{5}$Miller Institute for Basic Research in Science, University of California, Berkeley, CA, 94720, USA\\%
$^{6}$Physics Division, Lawrence Berkeley National Laboratory, Berkeley, CA 94720, USA\\%
$^{7}$Department of Physics and Astronomy, York University, 4700 Keele Street, Toronto, ON M3J 1P3, Canada\\%
$^{8}$Institute for Computational Cosmology, Department of Physics, Durham University, South Road, Durham, DH1 3LE, UK
}

\date{Accepted XXX. Received YYY; in original form ZZZ}
\pubyear{2022}

\begin{document}

\label{firstpage}

\pagerange{\pageref{firstpage}--\pageref{lastpage}}

\maketitle

\begin{abstract}
  We study weak gravitational lensing convergence maps produced from the {\sc MillenniumTNG} (MTNG) simulations by direct projection of the mass distribution on the past backwards lightcone of a fiducial observer. We explore the lensing maps over a large dynamic range in simulation mass and angular resolution, allowing us to establish a clear assessment of numerical convergence. By comparing full physics hydrodynamical simulations with corresponding dark-matter-only runs we quantify the impact of baryonic physics on the most important weak lensing statistics. Likewise, we predict the impact of massive neutrinos reliably far into the non-linear regime. We also demonstrate that the ``fixed \& paired'' variance suppression technique increases the statistical robustness of the simulation predictions on large scales not only for time slices but also for continuously output lightcone data. We find that both baryonic and neutrino effects substantially impact weak lensing shear measurements, with the latter dominating over the former on large angular scales. Thus, {\it both} effects must explicitly be included to obtain sufficiently accurate predictions for stage IV lensing surveys. Reassuringly, our results agree accurately with other simulation results where available, supporting the promise of simulation modeling for precision cosmology far into the non-linear regime. 
\end{abstract}


%%%%%%%%%%%%%%%%%%% INTRODUCTION %%%%%%%%%%%%%%%%%%%
\section{Introduction}
\label{sec:intro}

Cosmological  observations show that the majority of the present-day energy density of the Universe is composed of two mysterious ``dark'' components, with $\approx 70\%$ made up of Dark Energy and $\approx 25\%$ of Dark Matter; only the remaining $\approx 5 \%$ is baryonic. Understanding the physical nature of these two dark entities is one of the major challenges of modern cosmology. 

One particular cosmological probe that can help us to shed light on the dark sector is weak gravitational lensing~\citep[hereafter WL; for reviews see, e.g.,][]{Bartelmann2001,  Hoekstra2008, Kilbinger2015, Mandelbaum2018}. This effect has already shown its potential in constraining cosmological parameters with the results of Stage-III surveys like the KiDS \citep{Hildebrandt2016,Heymans2021}, DES \citep{Abbott2022} and HSC \citep{Aihara2022}. Upcoming Stage-IV WL surveys from Rubin \citep{LSST}, Euclid \citep{Amendola_2018}, and Roman \citep{Roman} will have higher resolution and larger sky coverage. They are poised to increase our knowledge of the dark  Universe substantially. In order to exploit the full potential of such surveys it is nevertheless necessary to have access to sufficiently accurate theoretical predictions for WL. 

% - Describe different WL simulations already present in literature (eg bahamas, ktng, takahashi):
Numerical simulations are the main tool for investigating the non-linear regime of WL. They offer a powerful way to identify potential systematics in observations. Even more importantly, simulations including the physics of baryons and/or massive neutrinos are required to understand how their effects impact the angular power spectrum and other WL observables. Motivated by this, several numerical methodologies for studying WL have been developed in the last decades. Among these are ray-tracing algorithms~\citep[e.g.][]{Hilbert2009}, the production of full-sky maps~\citep[e.g.][]{Fabbian2018,Hadzhiyska2023} and on-the-fly computation~\citep[e.g.][]{Barreira2016} of WL. Different numerical codes implementing these approaches have been compared in \citet{Hilbert2020}, which found them to produce consistent results provided certain resolution requirements are met. 

Upcoming observations will, however, require WL simulations with very high angular resolution that go beyond modeling based on CDM alone and on purely gravitational interactions. This is why there is now increasing interest in  WL predictions from high-fidelity cosmological simulations including additional components. Some recent papers along these lines include \citet{Osato2021}, \citet{Coulton2020} and \citet{Gouin2019}, which focus on the impact of baryons, as well as \citet{Fong2019} and \citet{Liu2018}  who study the impact of neutrinos. Their results suggest that both baryonic and neutrino effects should be included when interpreting data from upcoming stage-IV surveys.

% - Discuss the main WL observables (Pk, PDF, Peaks...). These can help braking degeneracies in the parameter space. Break the Sigma_8 - Omega_m degeneracy and inform the S_8 tension:

Previous work has also demonstrated that the WL signal contains important information beyond that in its two-point statistics~\citep[][]{Waerbeke2001,Bernardeau2002,Kilbinger2005}; this information can help break degeneracies in the cosmological parameter space, especially that between $\sigma_8$ and $\Omega_m$, thus shedding light on the so-called $S_8$ tension~\citep[see e.g.][]{Asgari2020}. Different higher-order observables have been considered, with popular examples including counts of peaks and minima in the convergence field~\citep[][]{Davies2022,Coulton2020,Fluri2018}, the one-point PDF of convergence~\citep[][]{Thiele2020, Liu2019}, three-point correlations~\citep[][]{Jung2021,Dodelson2005}, and Minkowski functionals~\citep[][]{Shirasaki2014, Kratochvil2012}. 

% - Introduce MTNG and what we are going to show in this work

In this paper we introduce our own method for computing high-resolution full-sky WL convergence maps, starting from the mass-shell outputs produced by our simulation set; we also present a way of efficiently partitioning a full-sky map into smaller square patches, which is based on the Fibonacci sphere distribution.  We apply our WL machinery to the {\sc MillenniumTNG} (MTNG) state-of-the-art simulation suite to study the impact of baryonic physics, massive neutrinos, and angular resolution. We also test how the use of fixed and paired initial conditions~\citep[see][]{Angulo2016} can improve the statistical robustness of WL simulations obtained from simulation boxes of limited size. The observables considered in this study are the angular power spectrum of the WL convergence, its one-point probability distribution function (PDF), and peaks and minima counts in the corresponding maps.

This study is part of the introductory paper set of the MTNG project. In \cite{Aguayo2022}, the technical aspects of the simulations are introduced together with a high-level analysis of matter and halo statistics. \cite{Pakmor2022} provide more detail of the hydrodynamical simulations, focussing, in particular, on the galaxy cluster population. \cite{Barrera2022} present an updated version of the {\sc L-Galaxies} semi-analytic modeling code and apply it to obtain lightcone output for the dark-matter-only simulations. \cite{Hadzhiyska2022a,Hadzhiyska2022b} present improved halo occupation distribution models for the halo--galaxy connection, focusing on the one-halo and two-halo terms, respectively. \cite{Bose2022} analyze galaxy clustering, in particular, as a function of the colour-selection. \cite{Delgado2022} investigate the intrinsic alignment of galaxy shapes and large-scale structure, and how it is affected by baryonic physics. \cite{Kannan2022} study the properties of the predicted galaxy population at $z>8$ in the full-hydro run. Finally, \cite{Contreras2022} shows how the cosmological parameters of MTNG can be recovered from mock SDSS-like galaxy samples, using an extended subhalo abundance matching technique combined with a fast-forward prediction model. 

This paper is organized as follows. In the remainder of this section, we introduce the mathematical formalism for weak lensing. In Section~\ref{sec:methods} we describe the methods we use to compute our WL maps and the associated observables. In particular, after giving an overview of the {\sc MillenniumTNG} simulation suite (Sec.~\ref{subsec:mtngproject}), we describe the ``mass-shell'' outputs (Sec.~\ref{subsec:massshell}) and how these are used in our code to produce WL convergence maps (Sec.~\ref{subsec:convmapsproduction}). We then introduce our method for partitioning a full-sky map efficiently into square patches via the Fibonacci sphere distribution (Sec.~\ref{subsec:patches}), and we briefly describe how the observables are extracted from the maps (Sec.~\ref{subsec:observables}). In Section~\ref{sec:results}, we begin by comparing results from  maps with different angular resolution (Sec.~\ref{subsec:resolution}). We then show the impact of baryonic effects (Sec.~\ref{subsec:baryons}) and massive neutrinos (Sec.~\ref{subsec:neutrinos}) on WL statistics. Lastly, we study the extent to which the use of fixed and paired initial conditions improves statistical robustness (Sec.~\ref{subsec:avsb}). In Section~\ref{sec:discussion} we compare our findings on the impact of baryons and massive neutrinos to  results from similar recent studies. In Section~\ref{sec:conclusions} we summarise our findings, concluding that WL lensing simulations aimed to inform stage-IV surveys must have high angular resolution and correctly model both baryonic and neutrino effects. In Appendix~\ref{appendix:smoothing} we show how different smoothing scales vary our results on the impact of resolution.

%------------Table----------------
\begin{table*}
\centering

\caption{Specifications of the simulations of the {\sc MillenniumTNG} project used in this paper.}

\begin{tabular}{cccccccccc}
\hline
Type      & Run name            & Series  &  Box size & $N_{\rm cdm}$	& $N_{\rm gas}$  & $N_{\nu}$ & Mass-shell $N_{\rm side}$ & $\sum m_{\nu}$ & $\epsilon_{\rm cdm}$  \\
	       &		             &	    	& $[\Mpch]$ & 		        &             &           &           & $[\rm eV]$ & $[\kpch]$             \\ \hline
DM only	  & MTNG740-DM-1	    & A/B	   & 500 &$4320^3$	    & $-$        &  $-$      & 12288    &  $-$  & 2.5                   \\
    	   & MTNG740-DM-2 	     & A/B		& 500 &$2160^3$	    & $-$         &  $-$      & 9182     &  $-$  & 5                     \\
    	   & MTNG740-DM-3	     & A/B		& 500 &$1080^3$	    & $-$         &  $-$      & 4096     &  $-$  & 10                    \\
	       & MTNG740-DM-4	     & A/B		& 500 &$540^3$	    & $-$         &  $-$      & 2048     &  $-$  & 20                    \\
	       & MTNG740-DM-5	     & A/B		& 500 &$270^3$	    & $-$         &  $-$      & 1024     &  $-$  & 40                    \\ \hline
Hydro 	  & MTNG740-1	        & A   	  &  500 &$4320^3$	    & $4320^3$  &  $-$      & 12288    &  $-$  & 2.5                   \\ \hline
Neutrinos & MTNG3000-DM-0.1$\nu$& A        & 2040&$10240^3$	    & $-$        &  $2560^3$	 & 12288    &  0.1  & 4                   \\
          & MTNG630-DM-0.3$\nu$ & A/B      & 430 &$2160^3$	    & $-$        &  $540^3$	 & 12288    &  0.3  & 4                   \\
    	   & MTNG630-DM-0.1$\nu$ & A/B		& 430 &$2160^3$	    & $-$         &  $540^3$   & 12288   &  0.1    & 4                    \\
	       & MTNG630-DM-0.0$\nu$ & A/B		& 430 &$2160^3$	    & $-$         &  $540^3$  & 12288    &  0.0   & 4                    \\ \hline
\end{tabular}

\label{tab:sims}
\end{table*}

\subsection{Weak gravitational lensing formalism}
\label{subsec:wlform}

During its travel from the source to the observer, light is deflected by the gravity of structures present along the path. Let us consider a Friedmann-Lemaître-Robertson-Walker universe with weak perturbations and denote with $\bm{\theta}$ and $\bm{\beta}$, respectively, the true and the observed position angles of a source which is located at a comoving line-of-sight distance $\chi_{\rm s}$ from the observer, corresponding to a redshift $z_{\rm s} = z(\chi_{\rm s})$. These two angles are related through the Newtonian gravitational potential $\Phi$ by means of the lens equation:
\begin{equation}
\label{eq:lens}
\bm{\beta}(\bm{\theta}, z_{\rm s})
 = \bm{\theta} - \frac{2}{{\rm c}^2} \int^{\chi_{\rm s}}_0 {\rm d} \chi_{\rm d} \frac{f_{\rm ds}}{f_{\rm d} f_{\rm s}} \nabla_{\bm{\beta}} \Phi (\bm{\beta}(\bm{\theta}, \chi_{\rm d}), \chi_{\rm d}, z_{\rm d}) \, ,
\end{equation}
where ${\rm c}$ is the speed of light, $f_{K}(\chi)$ is the comoving angular diameter distance related to the comoving line-of-sight distance $\chi$, and thus $f_{\rm ds} = f_{K}(\chi_{\rm s} - \chi_{\rm d})$, $f_{\rm d} = f_{K}(\chi_{\rm d})$ and $f_{\rm s} = f_{K}(\chi_{\rm s})$, and $\nabla_{\bm{\beta}}$ is the gradient with respect to the angular position $\bm{\beta}$. 
In the flat sky approximation, the Jacobian 
\begin{equation}
\label{eq:jacobian}
\frac{\partial \bm{\beta}}{\partial \bm{\theta}}
 = 
 \begin{pmatrix}
1 - \kappa - \gamma_1 & -\gamma_2 - \omega \\
-\gamma_2 + \omega    & 1 - \kappa + \gamma_1
\end{pmatrix}
\end{equation}
 can be written in terms of the lensing convergence $\kappa$, the lensing shear $\gamma = \gamma_1 + {\rm i}\gamma_2$, and the lensing rotation $\omega$. Equation~(\ref{eq:lens}) can be differentiated w.r.t.~to $\theta$, yielding:
\begin{equation}
\label{eq:lensder1}
\begin{split}
\frac{\partial \beta_i (\bm{\theta}, z_{\rm s})}{\partial \theta_j} = \delta_{ij} - & \frac{2}{{\rm c}^2} \int^{\chi_{\rm s}}_0 {\rm d} \chi_{\rm d} \frac{f_{\rm ds}}{f_{\rm d} f_{\rm s}} \\ 
& \times \frac{\partial^2 \Phi (\bm{\beta}(\bm{\theta}, \chi_{\rm d}), \chi_{\rm d}, z_{\rm d})}{\partial \beta_i \partial\beta_k} \frac{\partial \beta_k \bm{\beta}(\bm{\theta}, \chi_{\rm d})}{\partial \theta_j} \, ,
\end{split}
\end{equation}
where we introduced the Kronecker delta symbol $\delta_{ij}$. We now apply the Born approximation, that is we carry out the integral of the gradient of the potential along the unperturbed radial path $\bm{\theta}$, rather than along the actual light path, yielding:
\begin{equation}
\label{eq:lensder2}
\frac{\partial \beta_i (\bm{\theta}, z_{\rm s})}{\partial \theta_j} = \delta_{ij} - \frac{2}{{\rm c}^2} \int^{\chi_{\rm s}}_0 {\rm d} \chi_{\rm d} \frac{f_{\rm ds}}{f_{\rm d} f_{\rm s}} \frac{\partial^2 \Phi (\bm{\theta}, \chi_{\rm d}, z_{\rm d})}{\partial \theta_i \partial\theta_j} \, .
\end{equation}
This approximation is valid for the small deflections of light rays expected in the weak lensing regime.

We can now make use of the Poisson equation for the gravitational potential $\Phi$ and neglect boundary terms at the observer and source positions to obtain the following expression for the convergence:
\begin{equation}
\label{eq:kappa}
\kappa(\bm{\theta}, z_{\rm s})
 = \int^{\chi_{\rm s}}_0 {\rm d} \chi_{\rm d} \, q_{\rm ds} \, \delta_{\rm m} (\bm{\theta}, \chi_{\rm d}, z_{\rm d})
 \, ,
\end{equation}
where $\delta_{\rm m}$ is the density contrast and we introduced the lensing efficiency factor $q_{\rm ds}$, defined as:
\begin{equation}
\label{eq:lensfactor}
q_{\rm ds}= \frac{3 H_0^2 \Omega_m}{2 {\rm c}^2} (1+z_{\rm d}) f_{\rm d} \frac{f_{\rm ds}}{f_{\rm s}}
\, .
\end{equation}

By assuming statistical isotropy and applying a Limber-type approximation \citep{Limber1953, LoVerde2008}, one can furthermore obtain an equation that connects the angular power spectrum of the convergence $C_{\kappa} (\ell)$ to the three-dimensional matter power spectrum $P_{\rm m}$ ~\citep[see, e.g.,][for the complete derivation]{Hilbert2020}: 
\begin{equation}
\label{eq:pkmatter}
C_{\kappa}(\ell) = \int^{\chi_{\mathrm{lim}}}_0 \mathrm{d} \chi \frac{q^2_{\rm ds}}{f^2_{\rm d}} P_{\rm m} (\ell/\chi_{\rm d}, z_{\rm d}) \, .
\end{equation}
where $P_{\rm m} (\ell/\chi_{\rm d}, z_{\rm d})$ is the matter power spectrum evaluated at redshift $z_{\rm d}$ and wave number $k = \ell/\chi_{\rm d}$. This last equation shows how the convergence power spectrum mixes different 3D $k$-modes into 2D $\ell$-modes through the line-of-sight integration. It is possible to show that with the approximations made so far, i.e. Limber, flat-sky, and Born, the angular power spectra of the shear (E and B modes), convergence, and rotation, are related as follows:
\begin{subequations}
\label{eq:shearcl}
\begin{align}
  & C_{\gamma}^{\rm (EE)}(\ell) = C_{\kappa}(\ell) \, , \\
  & C_{\gamma}^{\rm (BB)}(\ell) = C_{\omega}(\ell) = 0 \, .  
\end{align}
\end{subequations}
Therefore in the present work, the WL convergence will be the only quantity taken into consideration.

%%%%%%%%%%%%%%%%%%% METHODS %%%%%%%%%%%%%%%%%%%

\section{Methods}
\label{sec:methods}

\subsection{The MTNG project}
\label{subsec:mtngproject}

The {\sc MillenniumTNG} (MTNG) project is based on a suite of high-resolution cosmological structure formation simulations. The project focuses on the connection between galaxy formation and large-scale structure
by combining the statistical power reached with the large box size of the  {\sc Millennium} simulation \citep{Springel2005}, with the high mass-resolution and sophisticated baryonic physics modeling of the {\sc IllustrisTNG} project \citep{Nelson2018, Springel2018, Marinacci2018, Pillepich2018, Naiman2018, Pillepich2019, Nelson2019a, Nelson2019b}. The goal of this synthesis, which inspired the name of the MTNG project, is to realize accurate and reliable theoretical predictions for galaxy formation throughout volumes large enough to be adequate for the upcoming surveys of cosmic large-scale structure.

The initial conditions of MTNG were generated at $z=63$ with an updated version of the {\sc N-GenIC} code, directly incorporated in {\sc Gadget-4}. The algorithm is based on second-order Lagrangian perturbation theory, and the input linear theory power spectrum was the same as the one used for the {\sc IllustrisTNG} simulations (based on  Planck15 cosmological parameters). A new transfer function with updated cosmological parameters was adopted for the simulations with massive neutrinos.

The dark matter (DM)-only simulations were run with the {\sc Gadget-4} code \citep{Springel2021}, using the variance-suppression technique introduced by \citet{Angulo2016}, so that for every resolution there are two simulations (which we refer to as A- and B-series) whose initial conditions are characterized by perturbations with opposite phases but the same amplitude, fixed to the {\it rms} value predicted by the power spectrum. The hydrodynamical simulations start from the same initial conditions as the DM A-series and were performed with the moving-mesh {\sc Arepo} code, featuring the same galaxy formation model as {\sc IllustrisTNG} \citep{Weinberger:2017MNRAS,Pillepich:2017jle}, modulo very small changes\footnote{Magnetic fields were not included, and the metallicity tracking was simplified. Both were necessary to reduce the memory consumption of the production run to make it fit into the available memory.}.

The main characteristics of the MTNG simulations that are primarily used in this work are summarised in Table~\ref{tab:sims}. 
For the bulk of our analysis, and in particular for studying the impact of resolution and of
baryons, we use a box size of $500\,h^{-1}{\rm Mpc} \simeq 740\,{\rm Mpc} $. For the
cosmological parameters, we use the \cite{Planck2015} cosmology,  which is consistent with what had been used for {\sc IllustrisTNG}: $\Omega_{\rm m} = \Omega_{\rm cdm} + \Omega_{\rm b} = 0.3089$, $\Omega_{\rm b} = 0.0486$, $\Omega_\Lambda = 0.6911$, $h = 0.6774$, $\sigma_8 = 0.8159$ and $n_s=0.9667$. For the study of the impact of massive neutrinos, we use a slightly smaller box size of $430\,h^{-1}{\rm Mpc} \simeq 630\,{\rm Mpc}$, and updated cosmological parameters that also take the different neutrino masses into account \citep{Abbott2022}. We consider three cases for the neutrino masses, $\Sigma m_\nu = 0$ meV (massless), 100 meV and 300 meV (see Table~\ref{tab:sims}).
The reader is referred to \citet{Aguayo2022} and \citet{Pakmor2022} for a more detailed description of the MTNG simulations. We simulate the effect of massive neutrinos using the so-called $\delta f$ method introduced in \citet{Elbers:2020lbn}, however, we refer to Hernández-Aguayo et al.~(2023, in prep) for a detailed description of the technical aspects of the simulations with neutrinos.

\subsection{Mass-shell outputs}
\label{subsec:massshell}

Along with snapshot and lightcone data \citep[see][for further details]{Aguayo2022}, the MTNG simulations provide ``mass-shell'' outputs, introduced as a new feature in {\sc Gadget-4}. These consist of a series of onion-shell-like full-sky maps built on-the-fly which store the line-of-sight projected matter field of the full-sky lightcone. Each shell consists of a HEALPix map \citep{healpix}, where each pixel contains, in turn, the total mass of all particles
that intersect the lightcone's time-variable hypersurface\footnote{This is simply the spherical surface whose comoving radius varies with time according to the finite propagation speed of light, and reaches the observer at the present time.} at a comoving distance that falls within the shell boundaries, and at an angular position that  falls within the solid angle corresponding to the pixel. For all the simulations of the MTNG suite, we fixed the comoving depth of these shells to $25\Mpch$. The angular resolution of a HEALPix map is modulated by the $N_{\rm side}$ parameter, which determines the total number of pixels through  $N_{\rm pix} = 12\,N_{\rm side}^2$. For simulations with increasing mass resolution, we typically constructed mass-shells with increasing $N_{\rm side}$, as can be seen in Table~\ref{tab:sims}. The highest angular resolution we reach with the mass-shells is $0.28$ arcmin, given by $N_{\rm side} = 12288$, which corresponds to approximately 1.8 billion pixels in the sky.

\subsection{Computation of full-sky convergence maps}
\label{subsec:convmapsproduction}

Starting from the mass-shell output, we developed a {\sc Gadget-4} post-processing python package for the computation of full-sky convergence maps in the Born approximation. This works as follows. The $i$-th mass shell can be converted into an angular surface mass density distribution $\Sigma$ dividing the mass at each pixel's angular position by the area of each pixel in steradians (given by $A_{\rm pix} = 4 \pi / N_{\rm pix}$, since HEALPix has equal area pixels):

\begin{equation}
\label{eq:surfmassdens}
\Sigma^{(i)}(\bm{\theta}) = \frac{M(\bm{\theta})}{A_{\mathrm{pix}}} \, .
\end{equation}

Every shell is then treated as a lens. For a fixed source redshift $z_s$, the convergence in the Born approximation will be given by integrating over the surface mass density at every lens plane (i.e. at every shell) between the source and the observer, weighted by the lensing efficiency factor:
\begin{equation}
\label{eq:efffactor}
\kappa(\bm{\theta}, \chi_s) = \frac{4 \pi \mathrm{G} }{\mathrm{c}^2} \frac{1}{f_{\rm s}} \sum_i (1+z^{(i)}_{\rm d}) \frac{f^{(i)}_{\rm ds}}{f^{(i)}_{\rm d}} \left[ \Sigma^{(i)}(\bm{\theta}) - \bar\Sigma^{(i)} \right] \, ,
\end{equation}
where $\bar\Sigma^{(i)}$ is the mean angular surface mass density of the $i$-th shell. In order to optimize computational efficiency, this calculation is parallelized with {\small MPI4PY} \citep{mpi4py}.

\subsection{Partitioning into square patches}
\label{subsec:patches}

\begin{figure}
\centering 
\includegraphics[width=41.5mm]{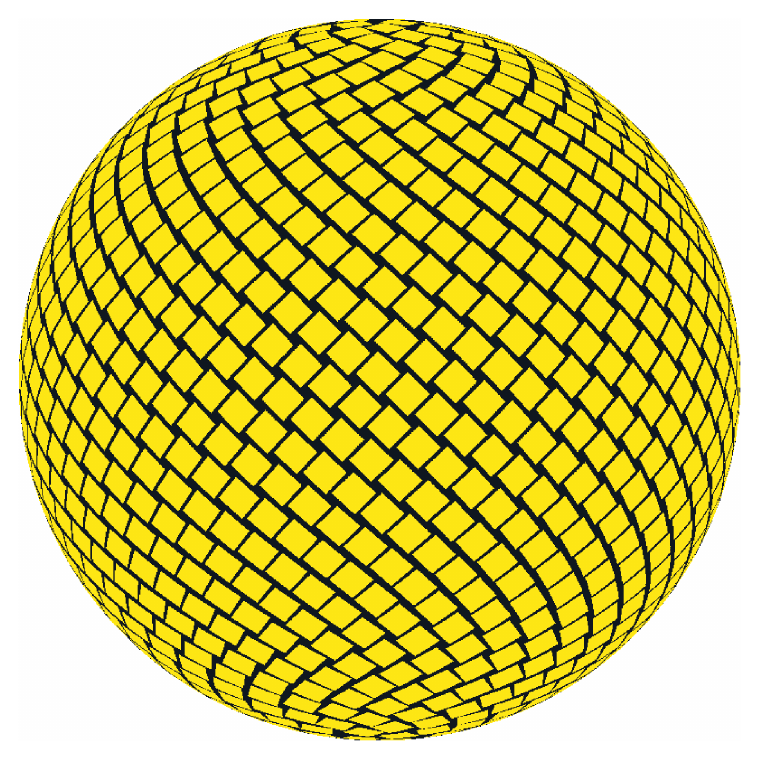}
\includegraphics[width=41.5mm]{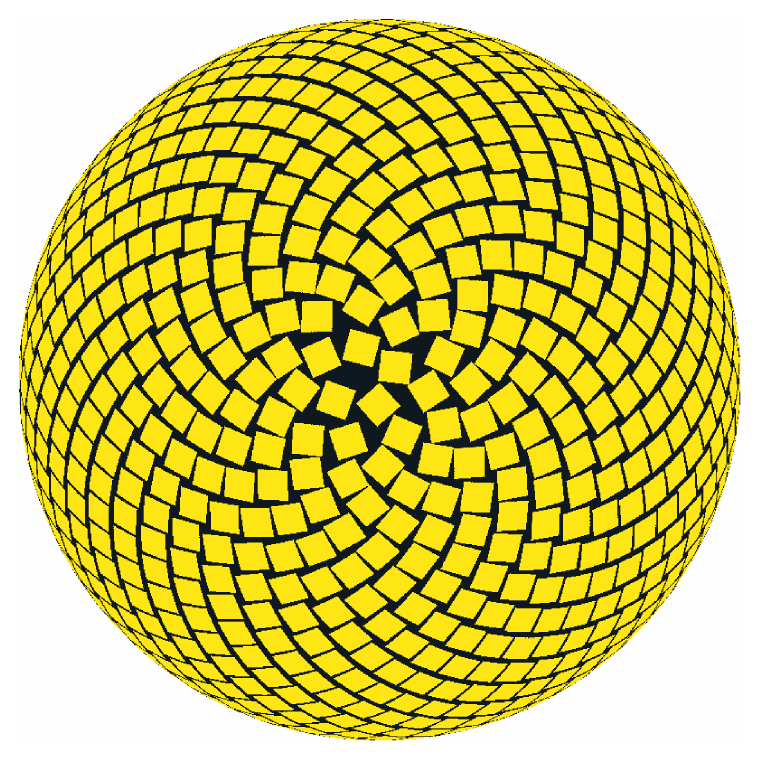}
\caption{Orthographic projection from the side (left) and from above (right) of the 1195 square maps with size $5 \times 5 \, \mathrm{deg}^2$ that we extract from a full-sky map. The method we use is based on the Fibonacci grid and manages to cover $\approx 72\%$ of the sphere surface.}\label{fig:fibonacci}
\end{figure}

\begin{figure}
    \centering
    \includegraphics[width=0.48\textwidth]{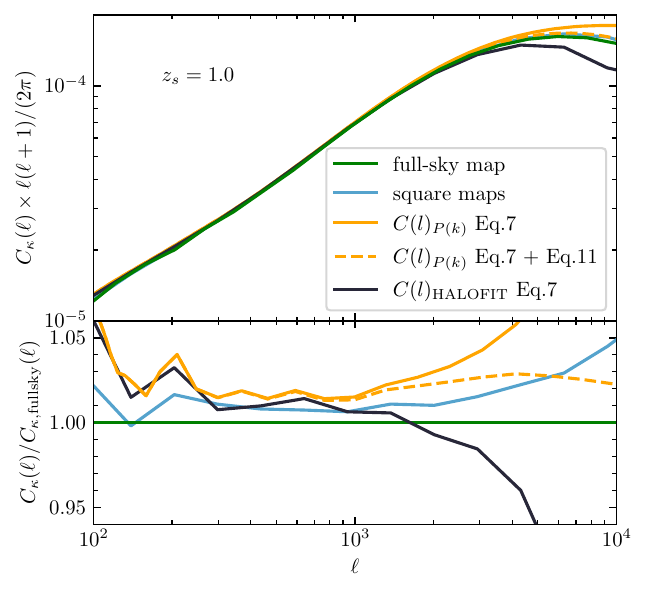}
    \caption{Lensing convergence power spectra (upper panel) of MTNG740-DM-1B obtained with our code from $N=1195$ non-overlapping $5 \times 5 \, \mathrm{deg}^2$ square maps (light blue line), and from the full-sky map (green line). These are compared with the convergence power spectrum obtained starting from the redshift-dependent non-linear 3D matter power spectra of the simulation (yellow lines, with the dashed line accounting for the suppression due to finite angular resolution as modeled by Eq.~(\ref{eq:dampfactor})) and from 3D power spectra given by the \textsc{Halofit} formula (black line). In both cases, the 3D matter power spectra are integrated according to Eq.~(\ref{eq:pkmatter}). Ratios relative to the full-sky map are shown in the lower panel.\label{fig:halofit}}
\end{figure}

\begin{figure*}
    \centering
    \includegraphics[width=1\textwidth]{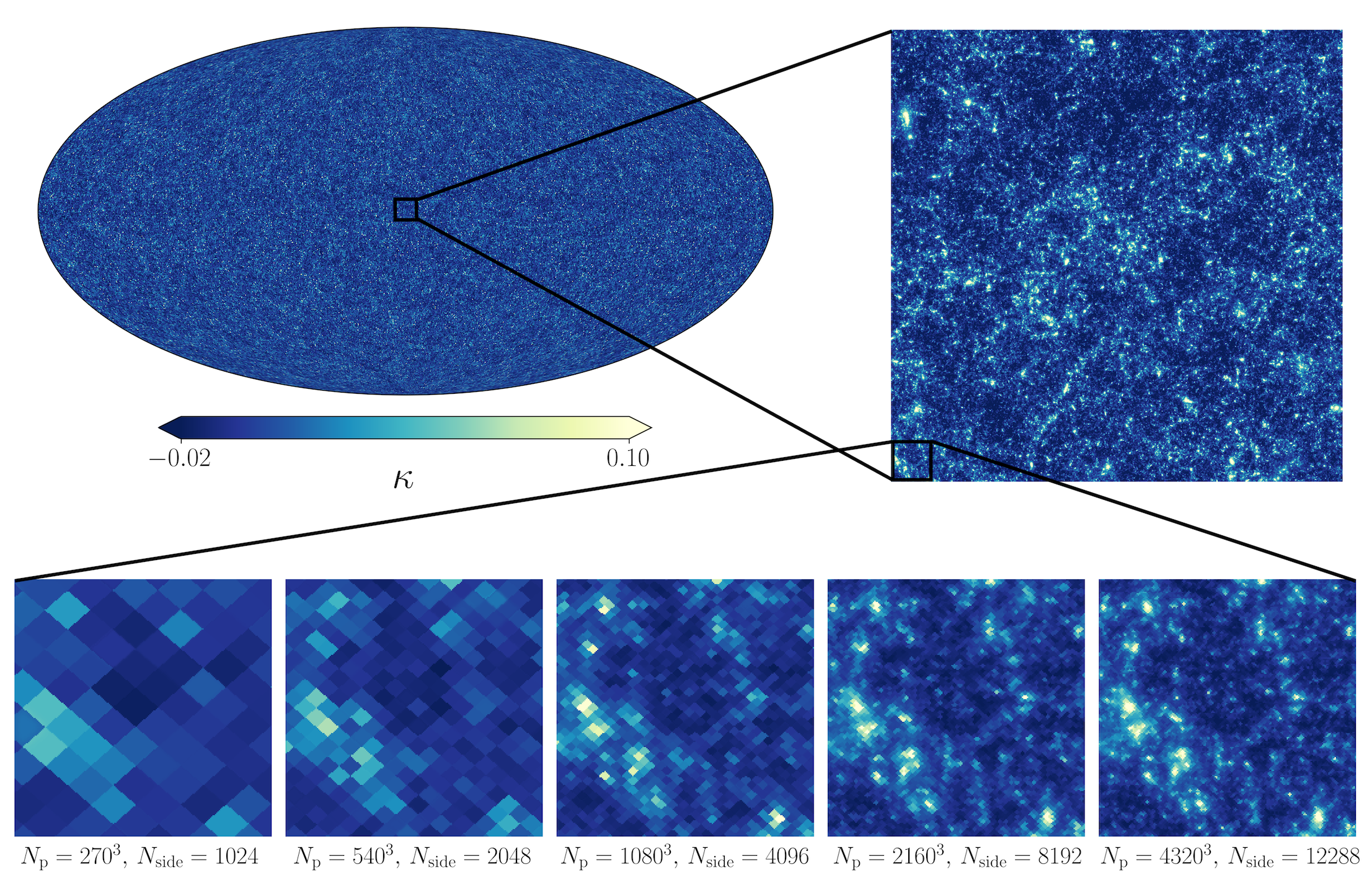}
    \caption{The top left shows a full-sky convergence map with $z_s = 1.0$ computed with our code from DM only runs with same initial conditions, but increasing resolution both in mass and in angles; the zoom on the top right focuses on a single $5 \times 5 \, \mathrm{deg}^2$ square patch. The bottom panels show a further zoom onto a square region of $0.5 \times 0.5 \, \mathrm{deg}^2$; these all represent the same region with increasing resolution from left to right.}\label{fig:zoom}
\end{figure*}

\begin{figure*}
    \centering
    \includegraphics[width=0.48\textwidth]{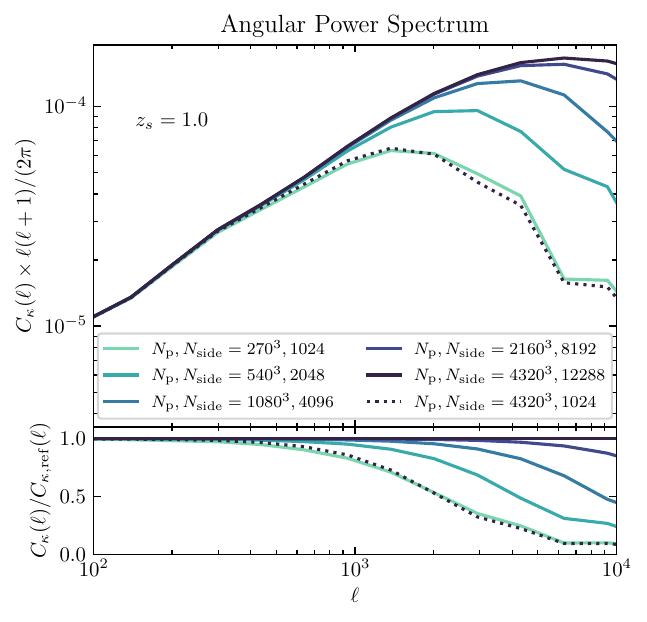}
    \includegraphics[width=0.48\textwidth]{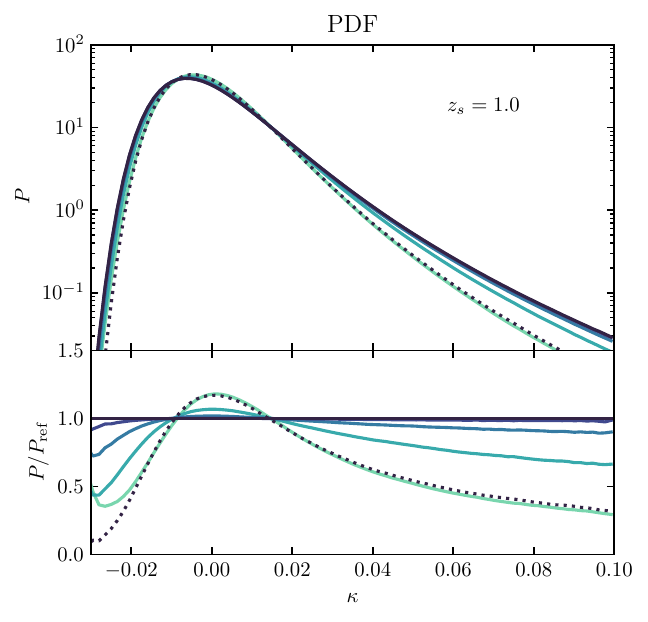}
    \includegraphics[width=0.48\textwidth]{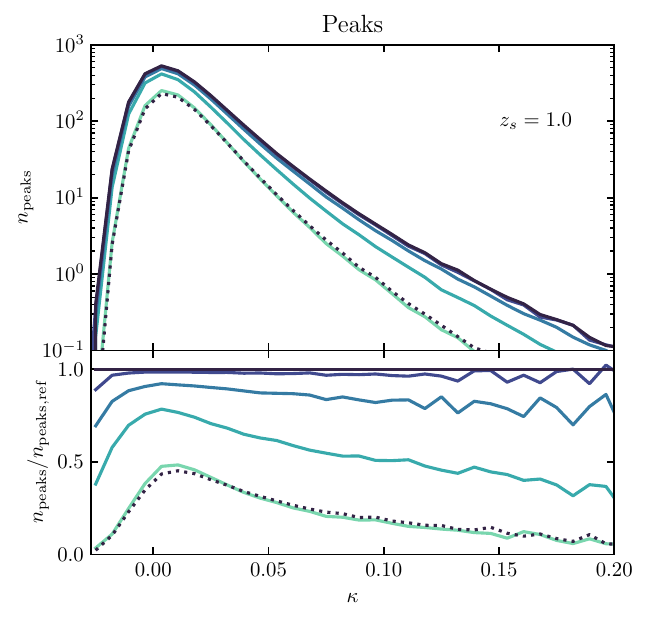}
    \includegraphics[width=0.48\textwidth]{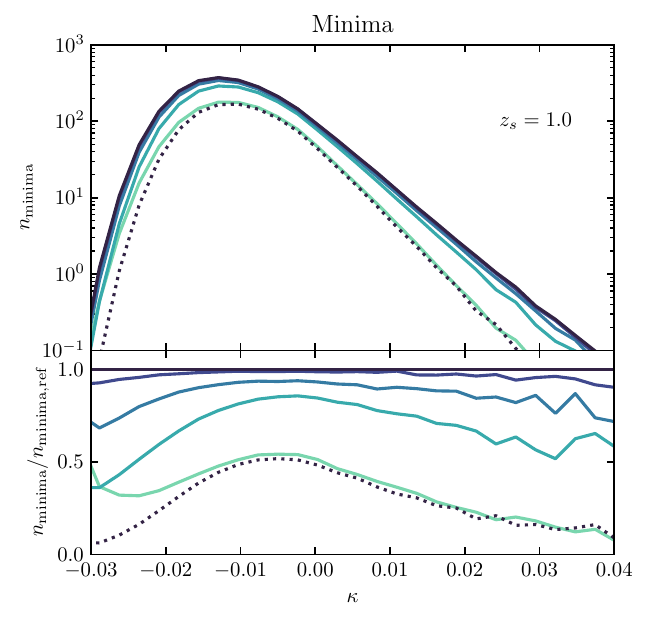}
    \caption{Top left: WL convergence power spectrum; top right: WL convergence PDF; bottom left: WL convergence peak counts; bottom right: WL convergence minimum counts. All the observables are computed on the B realization of the MTNG740-DM runs taking  $z_s = 1.0$. The solid lines indicate the mean of 1195 $5 \times 5 \, \mathrm{deg}^2$ square maps with increasing darkness representing increasing resolution both in mass and $N_{\mathrm{side}}$. The dotted line refers to the case with the highest mass resolution but down-sampled to $N_{\mathrm{side}}$ = 1024. In each lower sub-panel, we show the ratio w.r.t. the reference case with $N_{\mathrm{part}} = 4320^3$ and $N_{\mathrm{side}} = 12288$ (noted with the subscript "ref").}\label{fig:reso}
\end{figure*}

\begin{figure*}
    \centering
    \includegraphics[width=0.95\textwidth]{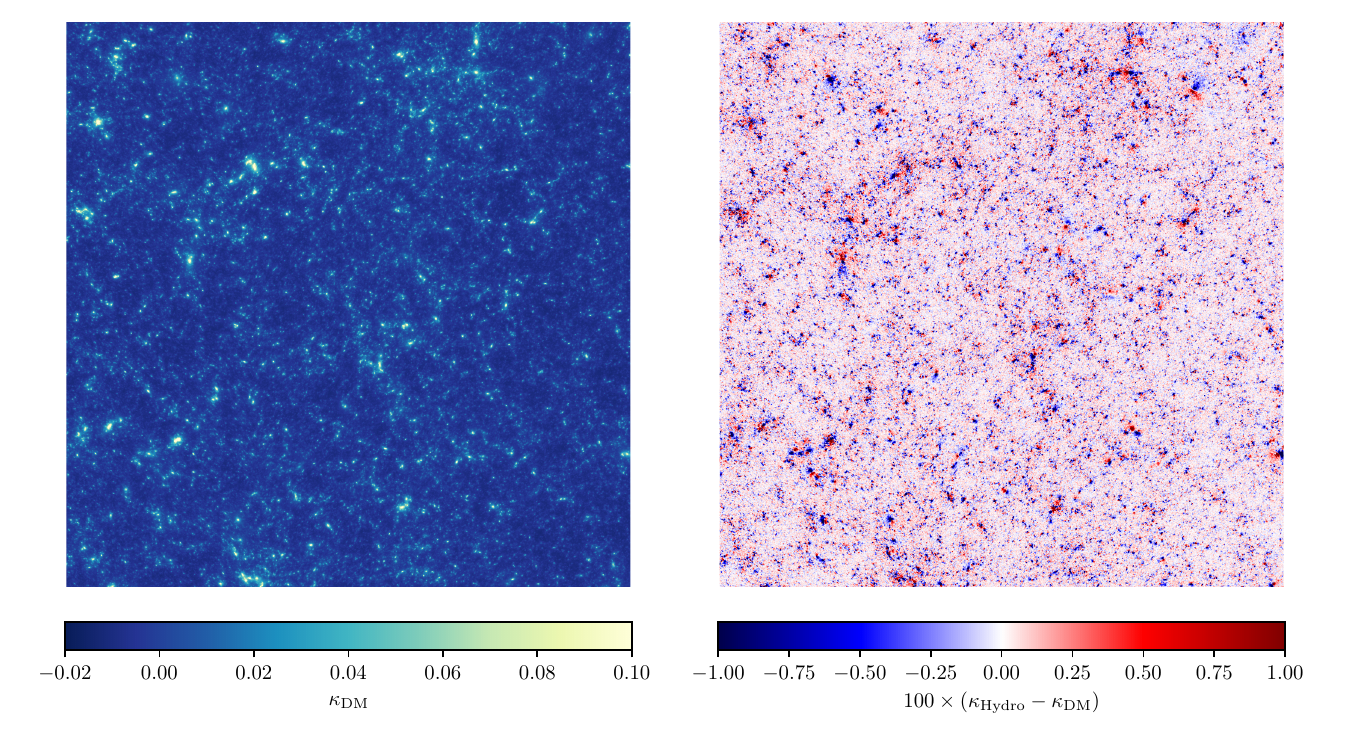}
    \caption{Left panel: one of our $5 \times 5 \, \mathrm{deg}^2$ square maps at $z_s = 0.5$ from MTNG DM-only run. Right panel: the same map but with the map from the corresponding MTNG hydro run (with the same initial conditions) subtracted.}\label{fig:map_compare}
\end{figure*}

\begin{figure}
    \centering
    \includegraphics[width=0.48\textwidth]{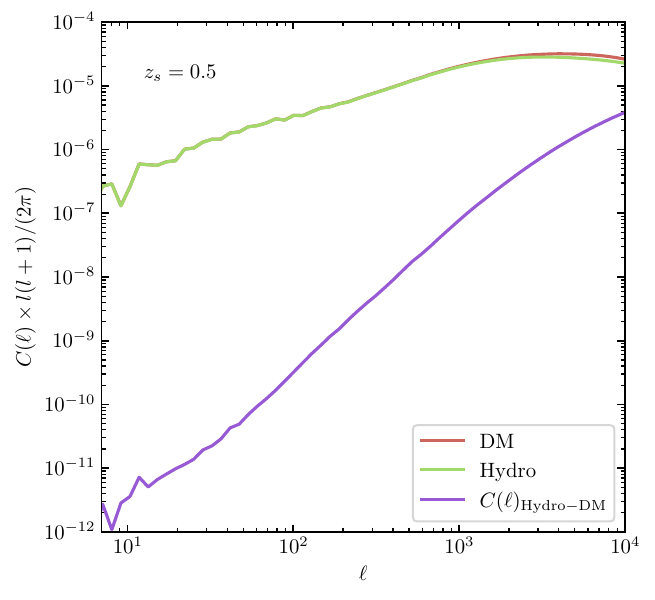}
    \caption{Power spectra of full-sky lensing convergence maps assuming $z_s = 0.5$. The red and green lines indicate results for the DM-only and Hydro runs, respectively. The purple line indicates the power spectrum of the difference between the two maps.}\label{fig:hydro_diff}
\end{figure}

\begin{figure*}
    \centering
    \includegraphics[width=0.99\textwidth]{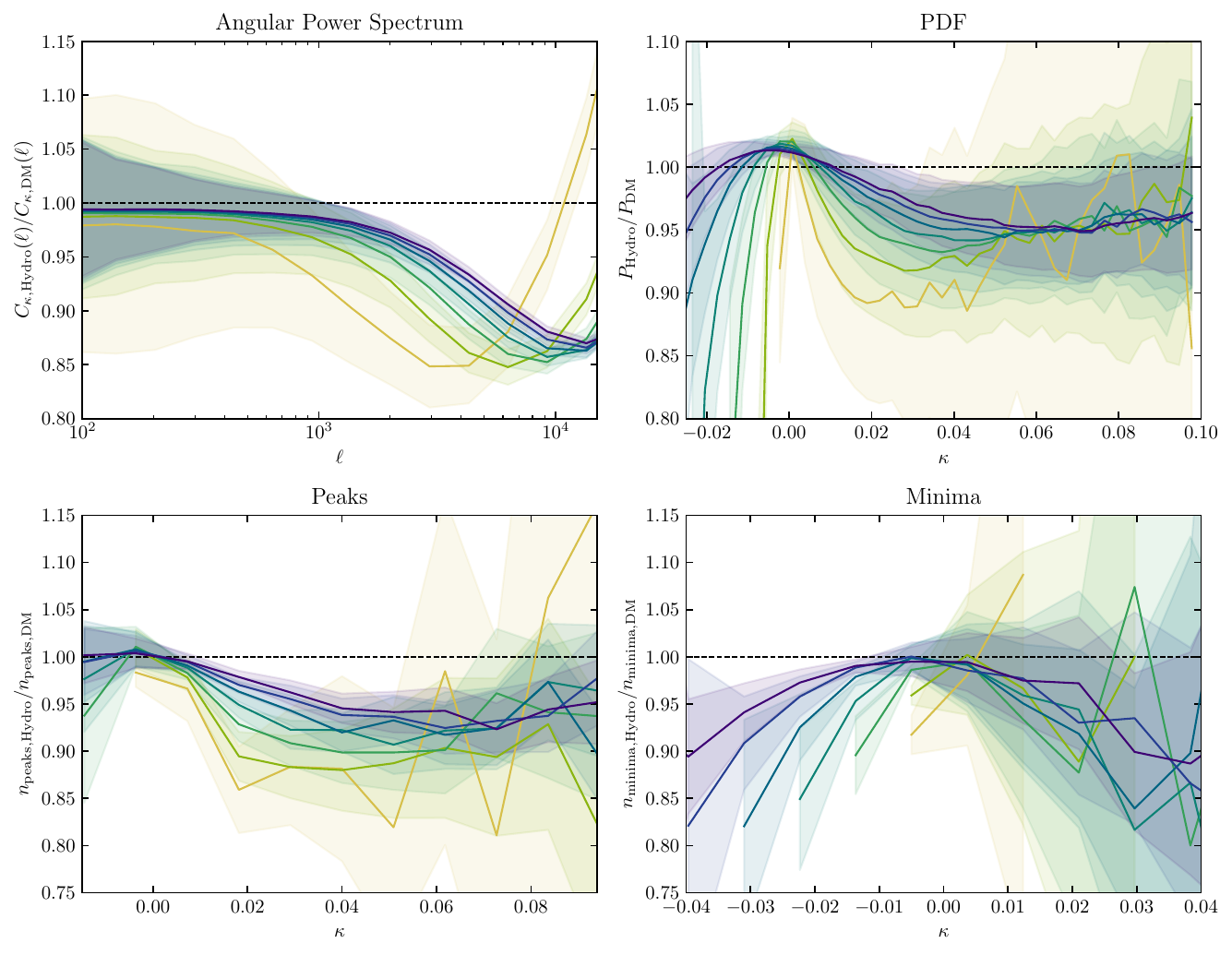}
    \includegraphics[width=0.47\textwidth]{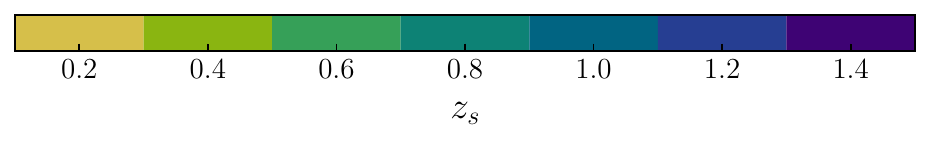}
    \caption{Top left: WL convergence power spectrum; top right: WL convergence PDF; bottom left: WL convergence peak counts; bottom right: WL convergence minimum counts. All the panels show the ratio of the results computed from full-hydro and DM-only runs obtained with our code considering  $z_s \in [0.2, 1.4]$ with $\Delta z_s = 0.2$. The solid lines indicate the mean of 125 $5 \times 5 \, \mathrm{deg}^2$ square maps, with the shaded regions representing the standard errors on the means.}\label{fig:hydro}
\end{figure*}

\begin{figure*}
    \centering
    \includegraphics[width=0.48\textwidth]{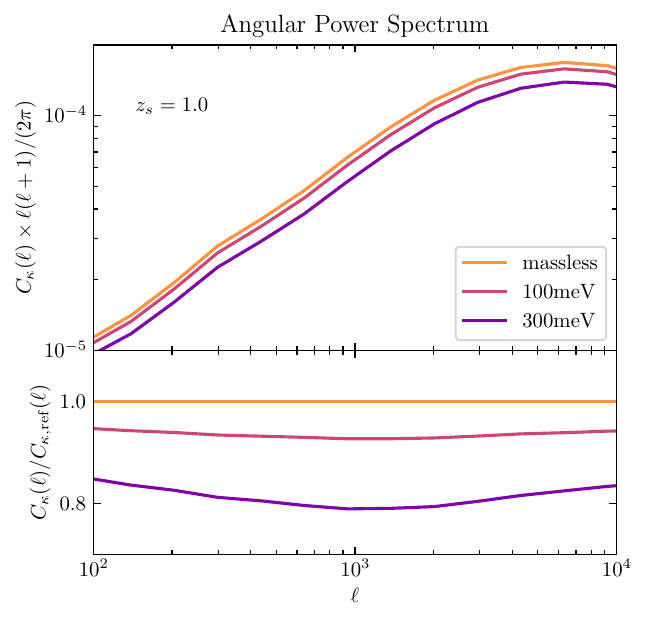}
    \includegraphics[width=0.48\textwidth]{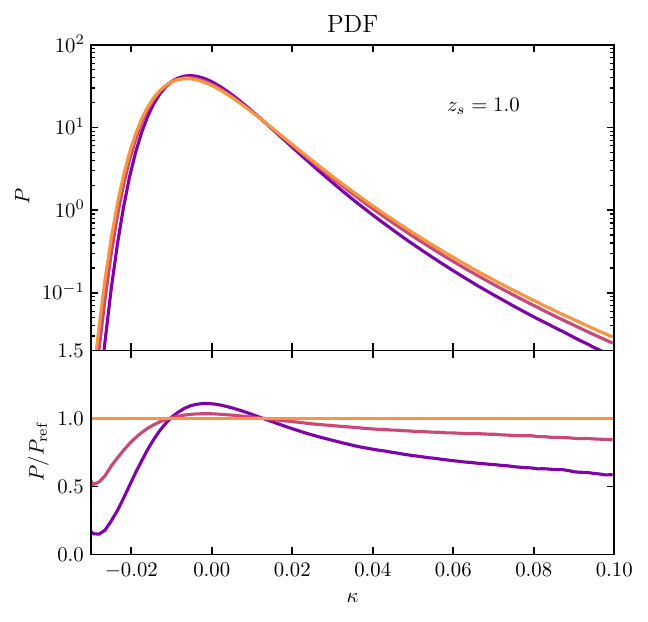}
    \includegraphics[width=0.48\textwidth]{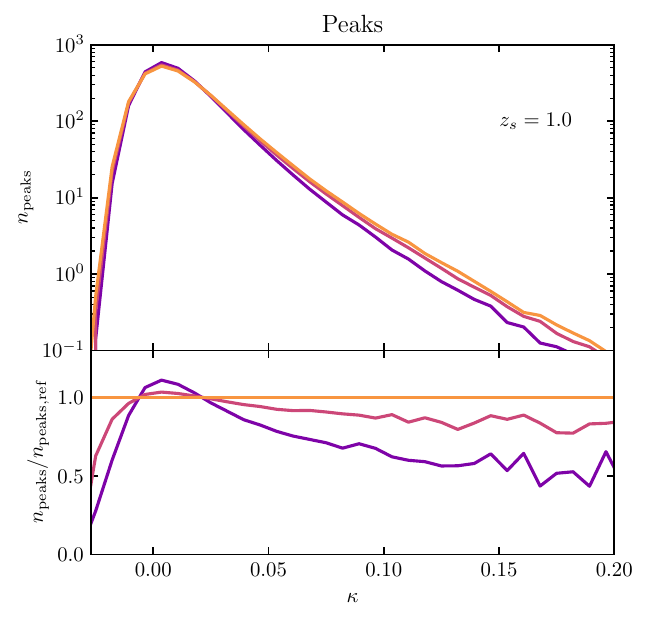}
    \includegraphics[width=0.48\textwidth]{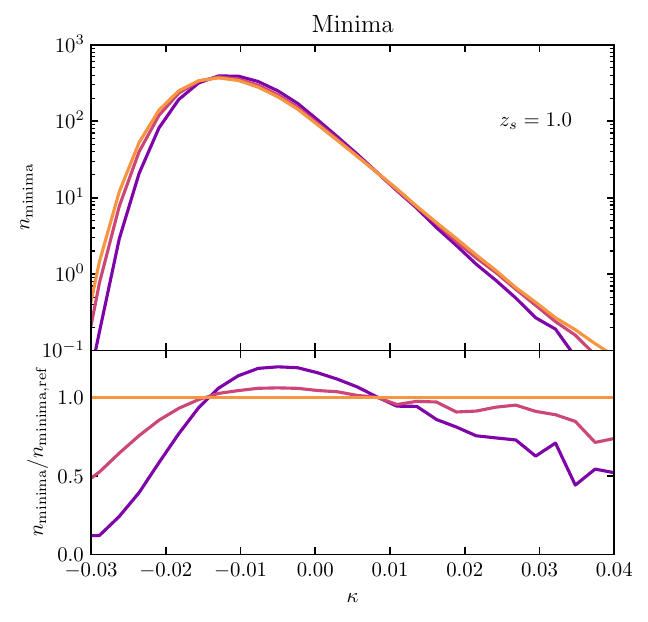}
    \caption{Top left: WL convergence power spectrum; top right: WL convergence PDF; bottom left: WL convergence peak counts; bottom right: WL convergence minimum counts. The orange, pink, and violet curves indicate the mean of 1195 $5 \times 5 \, \mathrm{deg}^2$ square maps computed on simulations with summed neutrino masses equal to $0, 100, 300 \rm \, meV $ respectively.The lower subpanels show the ratio of each distribution to that of the case with zero neutrino masses.}\label{fig:neutrinos}
\end{figure*}

Once a full-sky map is created, one may want to partition it into smaller non-overlapping square patches in order to simplify the analysis. Performing this operation in an efficient way, i.e.~covering as much as possible of the sphere's surface while avoiding overlap, is not a trivial task. An example is found in \citet{Davies2019}, where a HEALPix-based partitioning is performed to extract 192 maps with size $10 \times 10 \, \mathrm{deg}^2$; this scheme covers $\approx 47\%$ of the sphere's surface. 

In this work, we introduce a new and more efficient way of partitioning the sphere into smaller square maps. This is directly inspired by a botanical phenomenon known as \textit{phyllotaxis} (from Latin "leaf arrangement"), which refers to the way in which plants arrange their repeating parts (leaves, seeds, florets, etc...) in order to maximize the space occupation \citep[see e.g.][p.~113]{Conway_1996}. It turns out that in many cases (e.g. for the dandelion seeds or the florets on the sunflower head) the spatial distribution of points is mathematically described by the so-called Fibonacci grid. As shown in \citet{Swinbank_2006}, the spherical coordinates which describe the $i$-th point on a Fibonacci grid with a total of $2N + 1$ points are given by,
\begin{equation}
\label{eq:fibonacci}
\sin{\theta_i} = \frac{2i}{2N+1}, \,\,
\phi_i = \frac{2 \pi i}{\varphi}, \,\,
-N \leq i \leq N , \,\,
-\pi/2 \leq \theta_i \leq \pi/2,
\end{equation}
where $\varphi \approx 1.618$ is the golden ratio. We use these coordinates as the centers of our maps\footnote{Instead of following Eq.~(\ref{eq:fibonacci}), the first $\approx 10$ patches closest to each pole have been placed manually, in order to avoid overlapping.}. In addition, for square patches, we find that the coverage of the sphere is maximized when one diagonal of the squares lies on a meridian. Using this method, we place  $1195$ square patches of size $5 \times 5 \, \mathrm{deg}^2$; therefore covering $\approx 72\%$ of the sphere's solid angle (the same approximate percentage would also be reached in the case of $10 \times 10 \, \mathrm{deg}^2$ square patches).  The arrangement of spherical squares is shown in Figure ~\ref{fig:fibonacci}. Every square patch we extract is sampled on a regular grid with $2048^2$ pixels, resulting in a pixel size of about $ 0.14 \, \mathrm{arcmin}$. 

\subsection{Computation of the observables}
\label{subsec:observables}

We compute angular power spectra by means of the HEALPix \texttt{anafast} routine. This operation has been performed for maps with resolution up to $N_{\mathrm{side}} = 8192$, which marks the maximum resolution for which the HEALPix library is able to perform a spherical harmonics decomposition. In the case of square patches, the power spectra are calculated with Fourier transforms on a regular grid with $2048^2$ pixels in the flat-sky approximation, which is valid for the small field-of-view covered by their relatively small area ($5 \times 5 \, \mathrm{deg}^2$). The full-sky spectra are then binned into 80 equally spaced logarithmic bins in the range $\ell \in [10^0,10^4]$. The spectra extracted from the square patches are binned into 20 equally spaced logarithmic bins in the range $\ell \in [10^2,10^4]$. Before computing the probability distribution function for the convergence, and its peaks and minima statistics, all the square maps are smoothed with a Gaussian kernel characterized by a standard deviation of $1 \, \mathrm{arcmin}$. We compute the PDF in 50 linearly spaced convergence bins in the range $\kappa \in [-0.05,0.1]$. We identified peaks and minima as pixels in the maps that are greater or smaller than their 8 nearest neighboring pixels, respectively. We bin the peak counts into 50 equally spaced bins with $\kappa \in [-0.1,0.25]$, and the minima counts into 50 equally spaced bins with $\kappa \in [-0.07,0.06]$. We have fewer maps for the case that includes baryons, as explained in (Sec.~\ref{subsec:baryons}), therefore the peak counts are binned into 12 equally spaced bins with $\kappa \in [-0.02,0.1]$ and the minima counts into 16 equally spaced bins with $\kappa \in [-0.07,0.06]$. Statistics are computed using only one simulation run (B for the study on resolution, A for the impact of baryons and massive neutrinos) except in the section \ref{subsec:avsb} in which combined results from both A and B series runs are also calculated. Unless stated otherwise, all the observables are computed for a source redshift of $z_s = 1.0$. Finally, we do not include galaxy shape noise in this analysis, as the focus of this paper is to investigate the properties of the physical signal. 

%%%%%%%%%%%%%%%%%%% RESULTS %%%%%%%%%%%%%%%%%%%

\section{Results}
\label{sec:results}

We begin the presentation of our results with the following sanity check shown in Figure~\ref{fig:halofit}. We compute the convergence power spectrum in four different ways: 
\begin{itemize}
    \item We take the average of the convergence power spectrum computed on a large number of  $5 \times 5 \, \mathrm{deg}^2$ square maps extracted from the MTNG740-DM-1-A full-sky map.
    \item We compute the angular power spectrum of the full-sky map of the MTNG740-DM-1-A simulations by means of the HEALPix \texttt{anafast} routine.
    \item We use Eq.~(\ref{eq:pkmatter}) to obtain the convergence power spectrum by integrating over the 3D matter power spectra measured for MTNG740-DM-1-A  at the discrete set of snapshot times.
    \item Finally, we use the same approach but plug in the 3D matter power spectrum as
predicted by the \textsc{Halofit} emulation formula \citep{Halofit} using the CLASS code \citep{Blas2011}.
\end{itemize}
As Figure~\ref{fig:halofit} shows, we find quite good agreement between the four spectra. Those computed from  the full-sky map and from the square patches differ by less than $2.5\%$ over $100 \lesssim \ell \lesssim 4000$, indicating the validity of the flat sky approximation in this regime. The increasing discrepancy at smaller angular scales, i.e. for $\ell \gtrsim 4000$, is consistent with our predictions for different angular resolutions of the maps, as we will discuss in detail in the next section. The loss of power for \textsc{Halofit}  (black curve) at $\ell \gtrsim 4000$ is most likely  explained by the fact that this model was calibrated on simulations with lower resolution. On the other hand, we observe that the spectra computed from our maps (both full sky and square patches) tend to lose power at $\ell \gtrsim 4000$ with respect to the prediction obtained by plugging in the 3D matter power spectra computed from the simulation into Eq.~(\ref{eq:pkmatter}). This effect can be explained by the finite angular resolution of the maps. As noted by \citet{Takahashi2017}, one can approximately describe this effect by introducing a damping factor:
\begin{equation}
\label{eq:dampfactor}
C_{\kappa}(\ell) \rightarrow \frac{C_{\kappa}(\ell)}{1+(\ell/\ell_{\mathrm{res}})^2} \,\, .
\end{equation}
By setting the free parameter $\ell_{\mathrm{res}} = 2\, N_{\mathrm{side}}$ (yellow dashed curve), we recover the 5\% concordance at the smallest scales.

%%%%%%%%%%%%%%%%%% RESOLUTION %%%%%%%%%%%%%%%%%%%%%%%%%%%%%

\subsection{Numerical resolution study}
\label{subsec:resolution}

%%%%%%%%%%%%%%%%%% FULL-SKY ZOOM %%%%%%%%%%%%%%%%%%

In Figure~\ref{fig:zoom} we show an example of a full-sky convergence map computed with our code. We zoom in on such a map to show an example of an extracted $5 \times 5 \, \mathrm{deg}^2$ square patch. A further zoom-in to a $0.5 \times 0.5 \, \mathrm{deg}^2$ square region is performed to give a visual impression of how the different angular resolutions look at the smallest scales. By comparing this zoomed region for the $N_{\mathrm{side}}=1024, 2048$ and the $N_{\mathrm{side}}=8192, 12288$ cases, one can see that information on structures on the smallest angular scales is progressively lost as the resolution becomes lower. In this subsection, we investigate how this loss of angular resolution, in combination with reduced mass resolution in the simulation itself, can impact weak lensing observables extracted from the corresponding convergence maps. All the following observables are computed as averages of the 1195 square maps of size $5 \times 5 \, \mathrm{deg}^2$ extracted from the full-sky maps from the B realization of MTNG740-DM runs.

The first observable we study is the convergence power spectrum, as shown in the top left panel of Figure ~\ref{fig:reso}. The solid lines refer to simulations increasing both in mass resolution and angular resolution. This shows that decreasing the resolution (both in mass and angle) reduces the power at progressively larger scales. We verified that this effect is generally independent of $z_s$ over the range $z_s \in [0.2, 3.0]$. In order to understand how much of this reduction in power is due to the decrease in angular resolution and how much is due to the decrease in mass resolution, we now consider the convergence maps of the simulation with the highest mass resolution but down-sample the maps  to $N_{\mathrm{side}} = 1024$, this is represented by the dashed line. We see that for the fixed angular resolution $N_{\mathrm{side}} = 1024$ we obtain essentially the same result, independent of the mass resolution. The small-scale suppression we find with the decreasing resolution is consistent with the one found in previous studies \citep[see e.g.][]{Takahashi2017}.

Next, we consider how changes in the resolution impact the WL one-point PDF; the results are shown in the top right panel of Figure~\ref{fig:reso}. First, we see that the PDF is characterized by an asymmetrical shape that reflects the non-Gaussian nature of the WL convergence. The solid lines (which refer to increasing angular and mass resolution) show a broadening of the distribution when the resolution is increased. Since the maps with the higher angular resolution are able to capture the details of the smaller (angular) structures, more extreme convergence values are resolved; as seen in Figure~\ref{fig:zoom}, this results in a broader PDF. This explanation is supported by the dashed line (referring to the simulation with the highest mass resolution, but down-sampled to $N_{\mathrm{side}} = 1024$) which is almost indistinguishable from the solid line which refers to maps computed with the same angular resolution but from a simulation with much lower mass resolution. We find that angular resolution affects the convergence PDF similarly for source redshifts over the full range $0.2\leq z_s\leq 3$. The narrowing of the PDF we observe here is consistent with the suppression of the power spectrum seen previously. The comparison between these two indicates that the most extreme values of $\kappa$ are contained in the smallest angular scales.

Finally, we present that same investigation for the WL peak and minimum counts; the results are shown in the left and right lower panels of Figure~\ref{fig:reso}, respectively. For the peak counts, we find that when increasing resolution in angle and mass, the high $\kappa$ tail is more extended and the amplitude of the distribution increases. For the distribution of minimum counts also, we see an increase in amplitude with increasing resolution. In both cases, the suppression of the counts with decreasing resolution is not uniform; in particular, it is stronger for higher $\kappa$ values. Again, we conclude that the differences are dominated by the angular resolution, since the high-mass-resolution simulation, once down-sampled to $N_{\mathrm{side}} = 1024$, is again almost indistinguishable from the low-mass-resolution simulation analyzed with the same $N_{\mathrm{side}}$. As before, we find qualitatively similar results for source redshifts throughout the  range $z_s \in [0.2, 3.0]$. The results we find for peak and minimum abundance are consistent with what has been seen for the PDF: decreasing the resolution will narrow the PDF, therefore damping the most extreme values of $\kappa$, which in turn will result in fewer counts of peaks and minima. Finally, we tested the impact of varying the smoothing scale applied to the convergence map before the PDF, peak counts, and minima counts were computed, and found that the results are qualitatively consistent with Figure~\ref{fig:reso}. We refer the interested reader to Appendix~\ref{appendix:smoothing}, where we show and discuss these results.

%%%%%%%%%%%%%%%%%% BARYONS %%%%%%%%%%%%%%%%%%%%%%%%%%%%%

\subsection{Impact of baryons}
\label{subsec:baryons}

In this section, we study the impact of baryonic physics on weak lensing observables. In the left panel of Figure~\ref{fig:map_compare}, we show a $5 \times 5 \, \mathrm{deg}^2$ square patch with $z_s = 0.5$ extracted from a full-sky map for MTNG740-DM-1-A. We do not separately show the corresponding convergence map for the hydro run (which was run with the same initial conditions) as the difference with respect to the DM case is almost imperceptible by eye. Instead, in the right panel, we show the difference between the two maps. By comparing the two panels we notice that the regions where the baryonic physics has the strongest impact (redder and bluer areas in the right panel) roughly correspond to regions where the convergence map has high values (lighter areas in the left panel, corresponding to massive structures). We also see that the difference is often characterized by a dipole pattern (neighboring red-blue pairs). This largely reflects the fact that the same objects can end up having slightly different positions when baryonic physics is included and does not necessarily signal a significant difference in internal structure between the two cases.

In order to quantify how baryonic processes affect different angular scales, we compute the power spectrum of the difference between the two full-sky maps; this is shown as a purple line in Figure~\ref{fig:hydro_diff}, which is compared to the power spectra of the two individual maps. We see that the power spectrum of the difference map drops rapidly and approximately as a power law towards large scales.  The impact of baryonic physics increases strongly towards smaller angular scales over the range of $l$-values considered here.

We now consider results for the four primary observables considered previously, adopting a set of source redshifts over the range $z_s \in [0.2, 1.4]$ with $\Delta z_s = 0.2$. For the hydrodynamic simulation MTNG740-1, a  code configuration error, unfortunately, caused the loss of the original full-sky mass-shells for $z>0.5$, which were intended to be produced on-the-fly.  However, it proved possible to reconstruct these data partially in post-processing, because the full-particle lightcone of the simulation was stored for one octant of the sky out to $z=1.5$. Straightforwardly binning this data onto HEALPix arrays thus allows lensing maps to be recovered over 1/8-th of the full sky out to this redshift.  While this restricts us to a direct comparison of just 125 square maps (those that fall into the first octant), resulting in a somewhat larger statistical error (as indicated by the shaded regions that give standard errors), this does not substantially weaken our ability to assess the small-scale impact of baryonic physics.

In the top left panel of Figure~\ref{fig:hydro}, we show the ratio between the convergence power spectrum of DM-only and Hydro runs. We observe a small and almost constant suppression at the larger angular scales and a stronger, scale-dependent suppression at smaller scales. The transition between these two regimes takes place at $\ell \approx 10^3$ and happens at progressively larger $\ell$ with increasing $z_s$. The overall effect produces a spoon-shaped suppression which reaches $\approx 15 \%$. The dominant component of the power suppression can be explained in terms of feedback from black hole accretion and supernovae explosions, which blow away matter from the central regions of the halos. This will primarily  affect relatively small physical (and consequently angular) scales, but the associated redistribution of baryons induces an impact also on larger scales, particularly due to AGN as they are capable of affecting very massive halos. The shift of the spoon feature that we observe can be explained by considering that the physical scale at which the effect of baryons suppresses the most the power spectrum, will correspond to smaller angular scales (and therefore higher values of $\ell$) for increasing $z_s$.

We show the ratio between the WL PDF in the Hydro and in the DM-only cases in the top right panel of Figure~\ref{fig:hydro}. We find a roughly constant $\approx 5-10 \%$ suppression in the high-$\kappa$ tail for the Hydro run relative to the DM-only run. In the low-$\kappa$ regime, there is a suppression as well, and this increases dramatically as $\kappa$ becomes more negative. The central region of the PDF is in turn enhanced by $\approx 2-3 \%$. These changes impact a progressively broader $\kappa$ range as $z_s$ increases. Finally, we consider the effect of baryonic physics by considering the WL peaks and minima, shown in the right and left panels of Figure~\ref{fig:hydro},  respectively. In the case of the peak abundance, we observe a suppression of $\approx 5-15 \%$ for $\kappa \gtrsim 0.02$ which is stronger for decreasing $z_s$; although the results are noisier in the case of lower $z_s$. The distribution of the minima shows suppression in the baryonic case in both the high-$\kappa$ and low-$\kappa$ tails, and this effect increases the more $\kappa$ reaches extreme values. The trend is approximately symmetric and broader in $\kappa$ as $z_s$ increases. 

The effects we observe are consistent with the physical explanation given previously for the power spectrum: feedback processes redistribute matter from denser regions to lower-density regions. This manifests in a narrower PDF, and in a suppression of the peaks and minima counts. In particular, the high-$\kappa$ peaks are expected to correspond mostly to the presence of galaxy clusters along the line of sight, while the low-$\kappa$ peaks could be produced by haloes in voids or chance alignments of small haloes along the line of sight. We, therefore, expect the baryons to impact the peak abundance in a $\kappa$-dependent fashion. This could help in explaining the upturn we see for increasing $\kappa$ \citep[for a more detailed discussion we direct the reader to e.g.][]{Liu2016, Yang2011, White2002}.

Finally, we notice that the impact of baryonic physics on all the four observables is progressively stronger with decreasing $z_s$: we indeed expect this to happen because, at lower redshifts, baryonic processes have had more time to take place and therefore influence the overall cosmic structure.

%%%%%%%%%%%%%%%%%% NEUTRINOS %%%%%%%%%%%%%%%%%%%%%%%%%%%%%

\subsection{Impact of neutrinos}
\label{subsec:neutrinos}

Another important element that influences structure formation, and therefore WL observables, is the presence of massive neutrino species. In the early Universe, these act as an additional relativistic component thus delaying the onset of structure formation and suppressing the formation of structures below the free-streaming scale \citep[for reviews, the reader is redirected to e.g.][]{LeSourges2006, Won2011}. We, therefore, expect massive neutrinos to reduce the WL signal at those scales. Consequently, the use of WL has been suggested as a tool to constrain the neutrino mass \citep[see, e.g.,][]{Cooray1999aa}. In the following, we show results obtained by comparing MTNG DM-only with runs that include neutrino components with different overall mass contributions, corresponding to summed neutrino rest masses of $\sum m_{\nu} = [0, 100, 300] \,\mathrm{meV}$.

We start by considering the angular power spectrum, which is shown in the top left panel of Figure~\ref{fig:neutrinos}. We notice that this is suppressed by $\approx 5$ and $ 15 - 20 \%$ for $\sum m_{\nu} = 0.1$ and $0.3 \; \mathrm{eV}$, respectively, relative to the massless case. The suppression is slightly greater for intermediate angular scales $(\ell \approx 1000)$; this effect, which is barely noticeable in the $\sum m_{\nu} = 0.1\; \mathrm{eV}$ case, becomes more prominent for $\sum m_{\nu} = 0.3\; \mathrm{eV}$. Such an effect is consistent with massive neutrino species suppressing structure formation on small scales.  We verified that, as expected, this effect decreases significantly at the smallest $l$-values.

We show the convergence PDF in the top right panel of Fig.~\ref{fig:neutrinos}. Here the distribution is enhanced in its central region (for $ -0.015 \leq \kappa \leq 0.015$) of order $\approx 11$ and $ 4 \%$ for the $\sum m_{\nu} = 0.1$ and $0.3 \; \mathrm{eV}$ cases, respectively. On the other hand, we see that the PDF is progressively suppressed in the tails, an effect that gets stronger for higher neutrino masses. Interestingly, we find that the impact of massive neutrinos on the convergence PDF is quite similar to that induced by decreasing the angular resolution of the map, as one can notice by comparing this panel with the top left panel of Fig.~\ref{fig:reso}. The effect we observe is consistent with the physical interpretation given above: massive neutrinos will smooth out the density field, therefore narrowing the PDF of the WL convergence.

Finally, in the bottom panels of~Fig.~\ref{fig:neutrinos} we consider the effect of neutrinos on the convergence peak and minimum counts. For the peak counts, we see an enhancement for $ -0.015 \leq \kappa \leq 0.015$,  reaching $\approx 11 \%$ for $\sum m_{\nu} = 0.3 \; \mathrm{eV}$ and $\approx 4 \%$ for $\sum m_{\nu} = 0.1 \; \mathrm{eV}$. In the case of the minimum counts, the enhancement is present at $ -0.015 \leq \kappa \leq 0.01$ and reaches $\approx 20 \%$ and $ 6 \%$ for $\sum m_{\nu} = 0.1$ and $0.3 \; \mathrm{eV}$, respectively. Both the peak and minima counts are progressively suppressed along the tails of the distribution, and this effect becomes again stronger when the mass of neutrinos increases. What we observe is consistent with the effects on the PDF and the power spectrum. Massive neutrino species will tend to fill the emptiest regions, thus suppressing the negative-$\kappa$ tail of the minimum counts, and oppose to the formation of large structures, thus damping the high-$\kappa$ tail of peak counts.

%%%%%%%%%%%%%%%%%% A vs B %%%%%%%%%%%%%%%%%%%%%%%%%%%%%

\subsection{Paired and fixed initial conditions}

\begin{figure}
    \centering
    \includegraphics[width=0.48\textwidth]{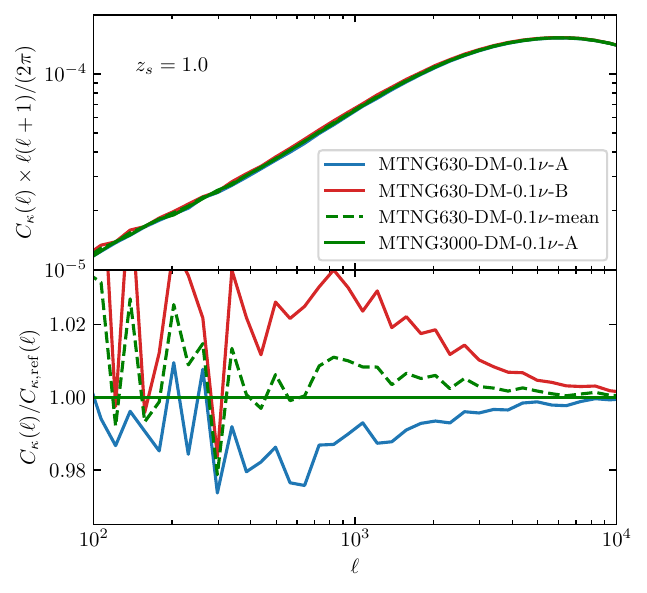}
    \caption{Lensing convergence power spectrum of full-sky convergence maps computed with our code considering $z_s = 1.0$. The blue and red lines indicate respectively the A and B series with fixed and paired initial conditions of MTNG630-DM-0.1$\nu$, the dashed green line refers to the mean of these two, while the solid green line to the run with the biggest box size, i.e. MTNG3000-DM-0.1$\nu$. In the lower sub-panel, we show the ratio w.r.t. the MTNG3000-DM-0.1$\nu$ run (noted with the subscript "ref").}\label{fig:pk_AB}
\end{figure}

\label{subsec:avsb}

To conclude the presentation of our primary results, we investigate the impact of the variance suppression technique introduced by \citet{Angulo2016}. As they show, averaging the 3D matter power spectra of two simulations with fixed and paired initial conditions can reduce the noise due to cosmic variance very significantly \citep[for another work that employs a variance suppression technique inspired by the previous citation, we direct the reader to][]{Harnois2019}. Here we test whether this approach also helps in the case of the convergence angular power spectrum. This requires a redshift integration over the lightcone, rather than a single time-slice of the underlying simulation, so it remains to be validated that the cancellation of second-order deviations from linear theory will work equally well in this case. 

We show our results for this in Figure~\ref{fig:pk_AB}, where the blue and red lines indicate the angular power spectra of full-sky convergence maps computed for the A and B versions of MTNG630-DM-0.1$\nu$, respectively, while the green dashed line shows their mean. For comparison, the power spectrum obtained from the full-sky lightcone of the A version of MTNG3000-DM-0.1$\nu$ is shown as a green solid line. Because of its larger box, the initial conditions of this simulation contain about 100 times as many modes on each scale as those of the smaller box simulations, and so the cosmic variance in its power spectrum is expected to be about 10 times smaller. We find that the power spectra of the smaller simulations differ from each other by up to 5\%  and from the power spectrum of the big simulation by up to 3\% for $300\leq l\leq 10^4$. Their mean, however, differs from the power spectrum of the big run by a maximum of 1\% and by much less at the smaller angular scales. Thus, although the suppression of cosmic variance is less strong than found by  \citet{Angulo2016} for the power spectra of the dark matter distribution in simulation snapshots, it is still very substantial, thus supporting the notion that the fixed and paired technique is an effective way to reduce cosmic variance uncertainties in simulation results also for WL observables.

%%%%%%%%%%%%%%%%%% DISCUSSION %%%%%%%%%%%%%%%%%%%%%%%%%%%%%

\section{Discussion}
\label{sec:discussion}

\begin{figure*}
    \centering
    \includegraphics[width=0.48\textwidth]{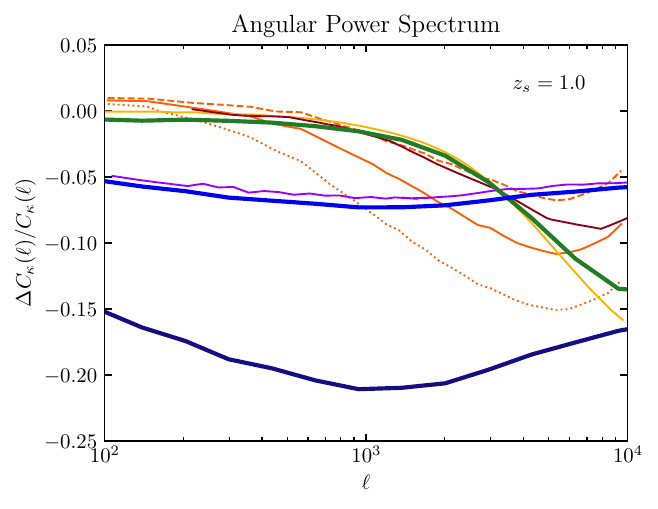}
    \includegraphics[width=0.48\textwidth]{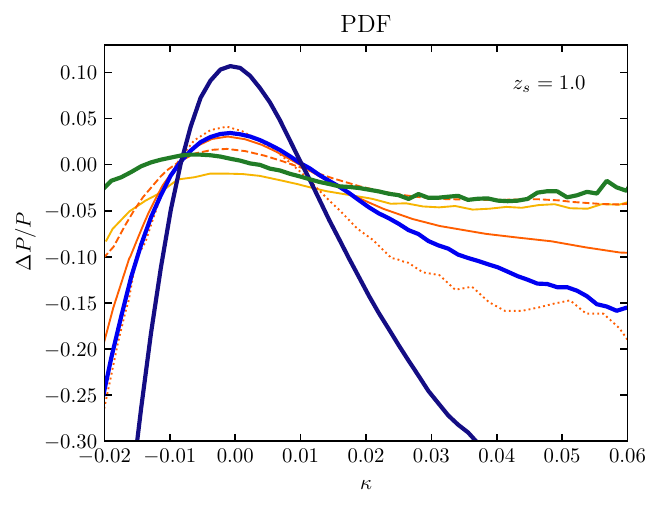}
    \includegraphics[width=0.48\textwidth]{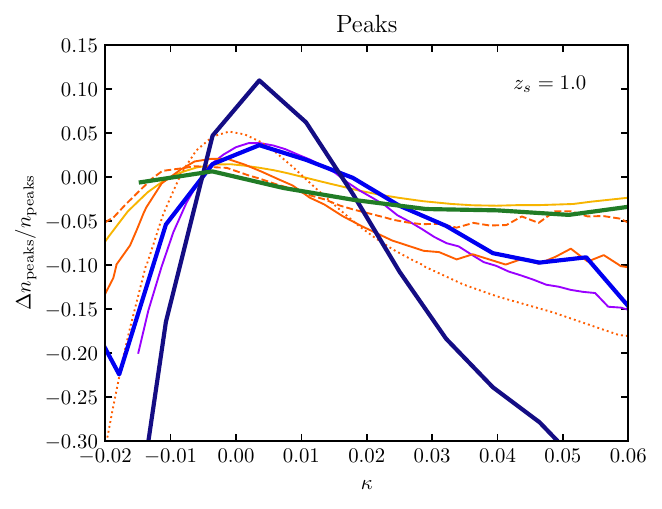}
    \includegraphics[width=0.48\textwidth]{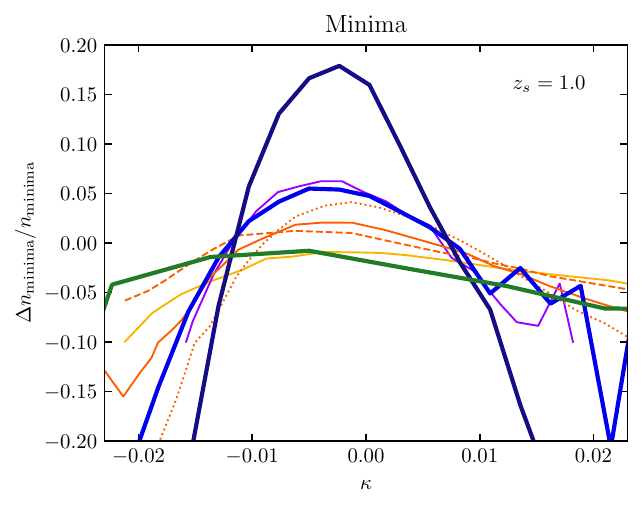}

\begin{tabular}{cccccccccc}
\hline
Color                                    & Simulation         & Code           & Box size  & $N_{\rm CDM}$ & $\sum m_{\nu}$ & $h$    & $\Omega_m$ & $\Omega_b$ & $\sigma_8$ \\
                                         &                    &                & $[\Mpch]$ &               & $[\rm eV]$     &        &            &            &            \\ \hline
\textcolor[HTML]{217C26}{$\blacksquare$} & MTNG740-1          & {\sc Arepo}    & 500       & $4320^3$      & -              & 0.6774 & 0.3089     & 0.0486     & 0.8159     \\
\textcolor[HTML]{0000F1}{$\blacksquare$} & MTNG630-DM-0.1     & {\sc Gadget-4} & 430       & $2160^3$      & 0.1            & 0.68   & 0.306      & 0.0487     & 0.8040     \\
\textcolor[HTML]{140D84}{$\blacksquare$} & MTNG630-DM-0.3     & {\sc Gadget-4} & 430       & $2160^3$      & 0.3            & 0.68   & 0.306      & 0.0487     & 0.7623     \\
\textcolor[HTML]{F7B500}{$\blacksquare$} & TNG300-1           & {\sc Arepo}    & 205       & $2500^3$      & -              & 0.6774 & 0.3089     & 0.0486     & 0.8159     \\
\textcolor[HTML]{FF5E00}{$\blacksquare$} & {\sc Bahamas}      & {\sc Gadget-3} & 400       & $1024^3$      & -              & 0.7    & 0.2793     & 0.0463     & 0.821      \\
\textcolor[HTML]{8D0116}{$\blacksquare$} & {\sc Horizon AGN}   & {\sc Ramses}   & 100       & $1024^3$      & -              & 0.704  & 0.272      & 0.045      & 0.81       \\
\textcolor[HTML]{9A00FF}{$\blacksquare$} & {\sc MassiveNuS}    & {\sc Gadget-2} & 512       & $1024^3$      & 0.1            & 0.7    & 0.3        & 0.046      & 0.8295     \\ \hline
\end{tabular}

    \caption{Top left: WL convergence power spectrum; top right: WL convergence PDF; bottom left: WL convergence peak counts; bottom right: WL convergence minimum counts. All the panels show the ratio between a simulation and its DM-only version. Here we compare our findings with those of similar recent studies. The results concerning baryonic effects are from {\sc MTNG} (green), {\sc TNG} (yellow), {\sc HorizonAGN} (brown), and {\sc Bahamas} (orange); {\sc Bahamas} comes in three different AGN intensities: low (dashed line), fiducial (solid line), and high (dotted line). The results concerning massive neutrinos are from {\sc MTNG} with $\sum m_{\nu} = 0.1 \; \mathrm{eV}$ (blue), {\sc MTNG} with $\sum m_{\nu} = 0.3 \; \mathrm{eV}$ (dark blue), and {\sc MassiveNuS} (purple).}\label{fig:compare}
\end{figure*}

We now discuss the implications of the results presented in the previous sections, in particular, for the relative impact of baryons and massive neutrinos on WL observations. Further, we compare our estimates of these effects with results from other recent studies.

\begin{table*}

\caption{Overview of the fractional differences between our results and those from other simulations for the four observables considered in this study: WL convergence power spectrum, PDF, peak counts, and minima counts. The values of the table are computed based on the results shown graphically in Figure~\ref{fig:compare}.}

\begin{tabular}{ccccccccc}
\hline
\multirow{2}{*}{Simulation} & \multicolumn{2}{c}{$C(\ell)$} & \multicolumn{2}{c}{PDF} & \multicolumn{2}{c}{Peaks} & \multicolumn{2}{c}{Minima}\\

                 & $\ell = 10^3$ & $ \ell = 10^4 $  & $\kappa=0$ & $\kappa=0.05$ & $\kappa=0$ & $\kappa=0.05$ & $\kappa=0$ & $\kappa=-0.02$  \\ \hline

\multicolumn{9}{c}{Baryons} \\

\hline
{\sc TNG300-1}         & $<0.01$       & 0.03             & 0.02       & 0.01          & 0.01       & 0.01          & $<0.01$    & 0.05            \\
{\sc Bahamas} low AGN  & $<0.01$       & 0.09             & 0.01       & $<0.01$       & $<0.01$    & $<0.01$       & 0.03       & 0.02            \\
{\sc Bahamas} fiducial & $ 0.02 $      & 0.05             & 0.02       & 0.05          & 0.02       & 0.04          & 0.04       & 0.10            \\
{\sc Bahamas} high AGN & $ 0.06 $      & 0.01             & 0.03       & 0.12          & 0.05       & 0.12          & 0.06       & 0.22            \\
{\sc Horizon AGN }     & $<0.01$       & 0.05             & /          & /             & /          & /             & /          & /               \\
\hline
\multicolumn{9}{c}{Massive neutrinos} \\
\hline

{\sc MassiveNuS}       & $<0.01$       & $<0.01$          & /          & /             & $<0.01$    & 0.03          & $<0.01$    & /               \\ \hline
\end{tabular}

\label{tab:compare}
\end{table*}

The four panels of Figure~\ref{fig:compare} show results for the four convergence map observables that we focus on in this paper, the angular power spectrum (top-left), the one-point PDF (top right), and the peak and minimum counts (bottom left and bottom right). In this case, in order to make the comparison consistent with other works, we smoothed the square maps with a Gaussian kernel with a standard deviation of $2 \, \mathrm{arcmin}$ when studying the PDF, peak, and minimum counts. As in Figure~\ref{fig:hydro} and in the lower subpanels of Figure~\ref{fig:neutrinos} we plot the ratios of results obtained in a simulation including either baryons or massive neutrinos to those for a simulation from identical initial conditions that followed only the CDM. First, we discuss the new results from this work, which is represented by the thicker lines: green for the baryons, blue for $\sum m_{\nu} = 0.1 \; \mathrm{eV}$, and dark blue for $\sum m_{\nu} = 0.3 \; \mathrm{eV}$. For the last case, neutrino effects dominate baryonic effects for all four observables: the suppression of the power spectrum, the distortion induced in the PDF, and the modification of the peak and minimum counts are all substantially stronger. On the other hand, for $\sum m_{\nu} = 0.1 \; \mathrm{eV}$ the baryonic and neutrino effects are comparable, though with different scale dependence, highlighting that these effects are partly degenerate. 

For the power spectra, the suppression induced by baryonic physics is negligible compared to that induced by neutrinos with $\sum m_{\nu} = 0.1 \; \mathrm{eV}$ for angular scales  $\ell < 2000$, but becomes dominant on smaller angular scales $\ell > 4000$. In the case of the PDF, peak counts, and minimum counts, we see that even a total neutrino mass of $\sum m_{\nu} = 0.1 \; \mathrm{eV}$ produces distortions that are larger than those induced by baryonic physics, especially in the tails of the distributions.

% discuss the consistency of our results with other studies

A crucial test to validate the reliability of results coming from numerical experiments is the comparison between independent, simulation studies. In Figure~\ref{fig:compare} results from previous studies are indicated by the thinner lines, yellow, brown, and orange referring to the baryonic effects computed for {\sc IllustrisTNG} \citep{Osato2021}, {\sc HorizonAGN}  \citep{Gouin2019}, and  {\sc Bahamas} \citep{Coulton2020},  respectively. This last simulation project considered three different models with increasingly strong  AGN feedback: these are indicated by dashed (\textit{low}) solid (\textit{fiducial}) and dotted (\textit{high}) lines. In purple, we include the result reported for massive neutrinos with $\sum m_{\nu} = 0.1 \; \mathrm{eV}$ for the {\sc MassiveNuS} simulations, where  the angular power spectrum effects were computed by \citet{Liu2019}, and the peak and minimum counts were computed in \citet{Coulton2020}.

Focusing first on the comparison with previous results for $\sum m_{\nu} = 0.1 \; \mathrm{eV}$ neutrinos, we see that our angular power spectrum modification agrees to $\approx 1 \%$  with that of {\sc MassiveNuS} at all angular scales. For the peak and minimum counts, we also find a reassuring $\approx 1-2 \%$ agreement for all values of $\kappa$ other than the high-$\kappa$ tails, where some statistical fluctuations are present. This exemplifies the robustness of such neutrino predictions, provided a sufficiently accurate simulation methodology is employed. This agrees with conclusions from the recent neutrino simulation code comparison project of \citet{Adamek2022}.

Next, considering the impact of baryons as predicted by different studies, we can directly compare the power spectrum modification we measure to results from the three independent projects included in Figure~\ref{fig:compare}. We find a $\approx 1-2 \%$ agreement  up to $\ell \approx 4000$ between MTNG and {\sc IllustrisTNG}, {\sc HorizonAGN} and the  `low AGN' variant of {\sc Bahamas}. At smaller angular scales, we see that {\sc MTNG} and {\sc TNG} predict a suppression that is of the same order as that in the high AGN variant of {\sc Bahamas}; for $\ell \approx 10^4$, this is stronger by roughly a factor of two than the predictions of {\sc HorizonAGN} and of the low and fiducial AGN versions of {\sc BAHAMAS}. Clearly the predictions are strongly affected by the specific implementation of feedback for $\ell > 10^3$. Even at $\ell = 10^3$ the fiducial version of {\sc Bahamas} predicts a 3\% effect, which is more than twice that found in the other models. 

Moving to the PDF, we find  $\approx 2 \%$ agreement between {\sc MTNG}, {\sc TNG}, and the {\sc Bahamas}'s low-AGN model, the only exception being for $\kappa \lesssim 0.01$, where our results predict a milder suppression of the tails. The fiducial and high AGN versions of {\sc Bahamas} deviate substantially, predicting a more extreme narrowing of the PDF. It is interesting to observe that these two cases become quite close to the effects seen for massive neutrinos with $\sum m_{\nu} = 0.1 \; \mathrm{eV}$. Similar conclusions can be drawn concerning the peak and minimum distributions, where we agree at $\approx 5 \%$ with  {\sc TNG}, and the low-AGN version of {\sc Bahamas}. Here also, the {\sc Bahamas}'s fiducial and (especially)  high AGN models differ substantially, and are quantitatively closer to the results for the $\sum m_{\nu} = 0.1 \; \mathrm{eV}$ neutrino case. As a summary of this discussion, we explicitly quantify the deviation between our results and the ones from other numerical studies in Table~\ref{tab:compare}.

Overall, our results reaffirm the strong sensitivity of WL observables to AGN feedback physics, and they also point out an important partial degeneracy between the impacts of massive neutrinos and baryonic physics. At the same time, the comparatively good agreement between different simulation methodologies can be seen as encouraging, underlining the predictive power of the simulations far into the non-linear regime, even though the high systematic uncertainty in the strength of AGN feedback clearly remains a point of concern.

%%%%%%%%%%%%%%%%%% CONCLUSIONS %%%%%%%%%%%%%%%%%%%%%%%%%%%%%

\section{Conclusions and outlook}
\label{sec:conclusions}

In this paper, we introduced our methodology for computing full-sky maps of  weak lensing convergence, starting from the mass-shell outputs of {\sc Gadget-4}. We applied our code to a selection of simulations from the {\sc MillenniumTNG} suite, presenting predictions for four observables, namely the angular power spectrum, the one-point PDF, and counts of peaks and minima as a function of convergence. 

After assessing the internal consistency of our code by comparing our results to theoretical predictions, we investigated the impact of mass and angular resolution on the weak lensing convergence, finding low angular resolution to be particularly problematic. Even if the underlying simulation has high mass resolution, insufficient angular resolution causes significant suppression of the angular power spectrum at small scales, as well as a narrowing of the one-point PDF and an underprediction of the numbers of peaks and minima at all values of the WL convergence. Creating convergence maps featuring high angular resolution is therefore of critical importance, arguably even more important than the underlying mass resolution. We also tested whether the ``fixed and paired'' variance-suppression technique proposed by \citet{Angulo2016}  remains beneficial when applied to continuous lightcone output over a wide redshift range, rather than to individual  simulation snapshots. We found that it does indeed significantly reduce cosmic variance uncertainties in the angular power spectrum of WL convergence at medium to large $\ell$-values.

We investigated the impact of baryonic physics on WL measurements by comparing convergence maps from DM-only and full-hydro simulations run from identical initial conditions. We found that including the baryons results in a redshift-dependent suppression of angular power which can reach $\approx 15 \%$ for $\ell \gtrsim 10^3$. The PDF in turn becomes narrower, increasingly so for higher source redshift, while the counts of peaks and minima are suppressed in an $\kappa$-dependent fashion by up to $\approx 15 \%$. This emphasizes the need to include the impact of baryons in any attempt to model WL observables with high precision.

We also studied the effect of massive neutrinos on WL observables by comparing simulations with different total neutrino masses, viz.~$\sum m_{\nu} = [0, 100, 300] \,\mathrm{meV}$. The impact is significant, and  is especially drastic for $\sum m_{\nu} = 300 \,\mathrm{meV}$, producing a suppression of the angular power spectrum by up to $\approx 20\%$, and inducing a significant distortion of the PDF and of the distributions of peak and minimum counts, primarily a suppression of the tails and an enhancement of the central parts of the distributions.

In summary, weak lensing predictions of the precision needed to interpret stage IV surveys {\it require} appropriate modeling of the impact {\it both} of baryonic physics and of massive neutrinos. Furthermore, these must be implemented in simulations which simultaneously have {\it both} sufficiently high angular and mass resolution {\it and} large enough periodic box size.  In the present study we adopted a purely theoretical perspective, focusing exclusively on the mass distributions predicted by the MTNG simulations.  However, this simulation suite also predicts the properties of the galaxies themselves, either directly in the large hydrodynamical simulation MTNG740, or through semi-analytic modeling throughout the extremely large-volume  DM-only simulation MTNG3000 (which also includes massive neutrinos). Forthcoming work will thus consider realistic forward modeling of weak lensing observations in order to study various correlations between the WL signals and the galaxy distribution.

\section*{Acknowledgements}

We thank the referee for a constructive and insightful report that helped to improve this paper.
FF would like to thank Martin Reinecke, Soumya Shreeram, and Nicola Locatelli for useful comments and discussions. VS and LH acknowledge support by the Simons Collaboration on ``Learning the Universe''. 
SB is supported by the UK Research and Innovation (UKRI) Future Leaders Fellowship [grant number MR/V023381/1]. 
LH is supported by NSF grant AST-1815978.  
CH-A acknowledges support from the Excellence Cluster ORIGINS which is funded by the Deutsche Forschungsgemeinschaft (DFG, German Research Foundation) under Germany's Excellence Strategy -- EXC-2094 -- 390783311. 
The authors gratefully acknowledge the Gauss Centre for Supercomputing (GCS) for providing computing time on the GCS Supercomputer SuperMUC-NG at the Leibniz Supercomputing Centre (LRZ) in Garching, Germany, under project pn34mo. 
This work used the DiRAC@Durham facility managed by the Institute for Computational Cosmology on behalf of the STFC DiRAC HPC Facility, with equipment funded by BEIS capital funding via STFC capital grants ST/K00042X/1, ST/P002293/1, ST/R002371/1 and ST/S002502/1, Durham University and STFC operations grant ST/R000832/1. 

\section*{Data Availability}
The simulations of the {\sc MillenniumTNG} project will be made fully publicly available in 2024 at the following address: \url{https://www.mtng-project.org}. The data underlying this article will be shared upon reasonable request to the corresponding authors.

\bibliographystyle{mnras}
\bibliography{MTNG_WL}

\appendix

\section{Impact of different smoothing scales and resolutions}
\label{appendix:smoothing}

\begin{figure*}
    \centering
    \includegraphics[width=0.325\textwidth]{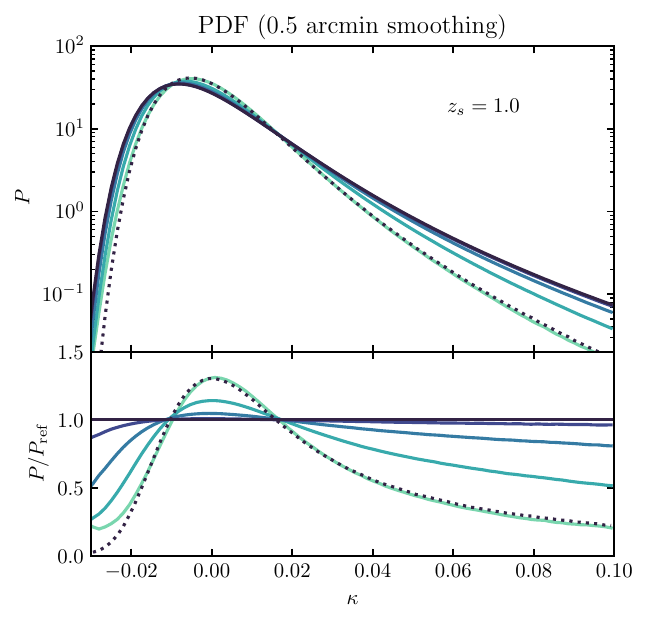}
    \includegraphics[width=0.325\textwidth]{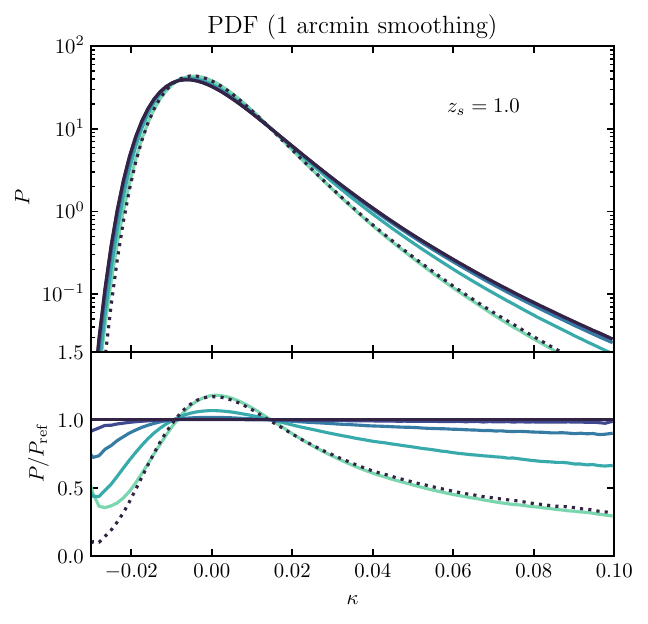}
    \includegraphics[width=0.325\textwidth]{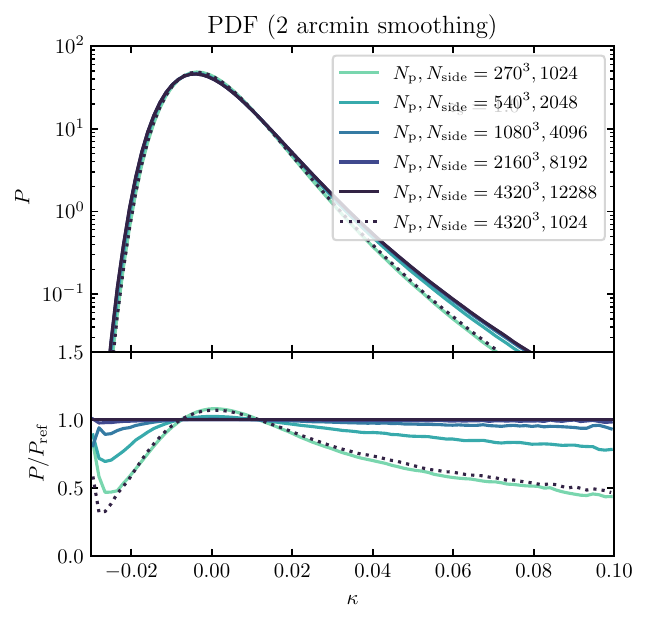}
    \includegraphics[width=0.325\textwidth]{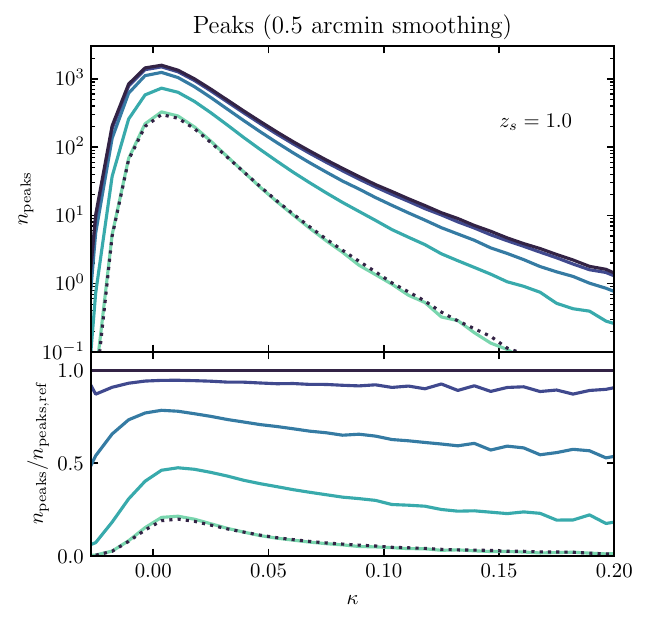}
    \includegraphics[width=0.325\textwidth]{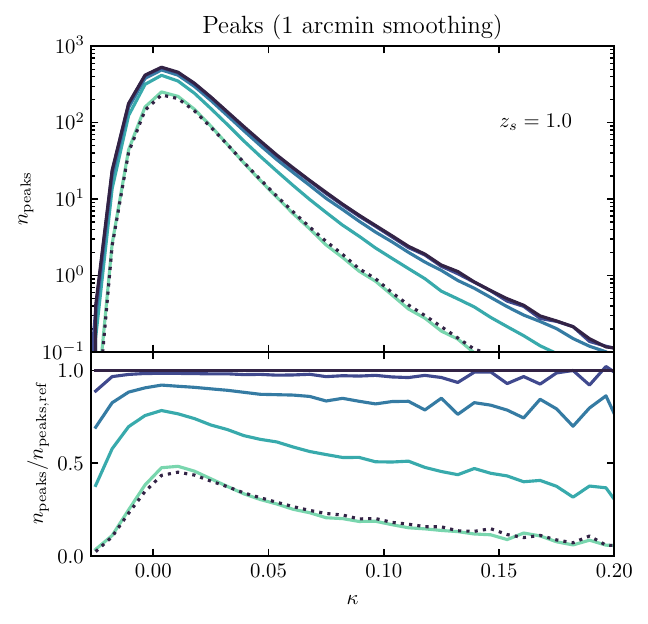}
    \includegraphics[width=0.325\textwidth]{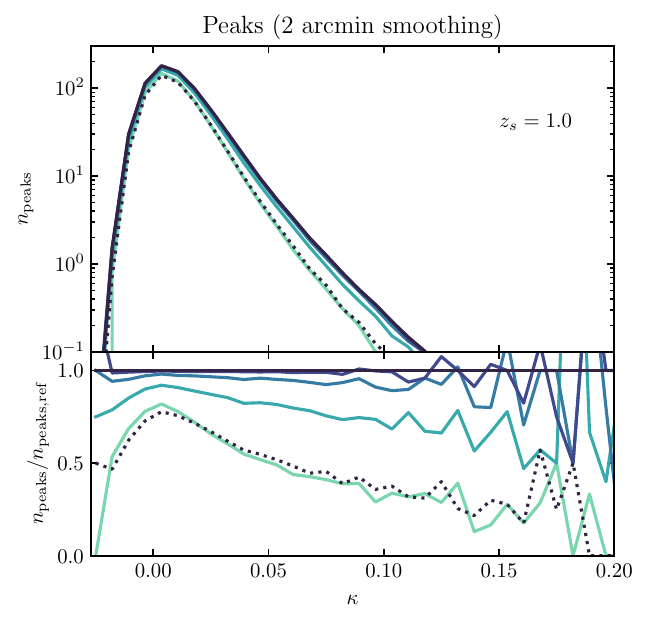}
    \includegraphics[width=0.325\textwidth]{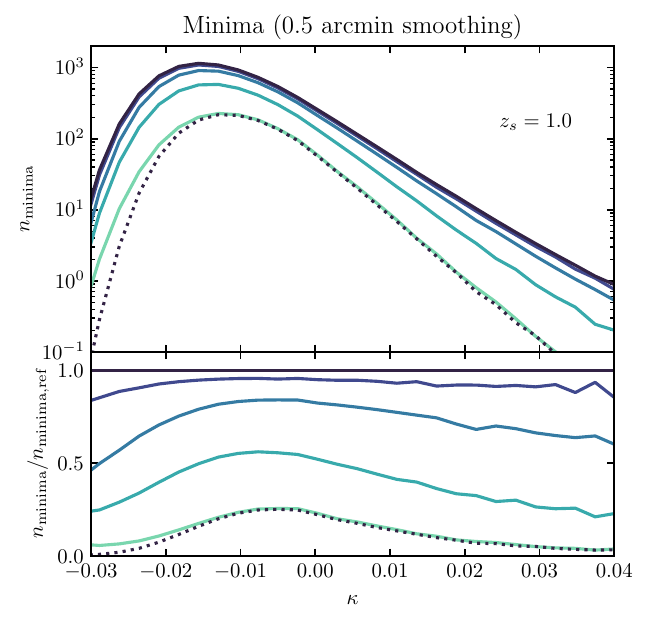}   
    \includegraphics[width=0.325\textwidth]{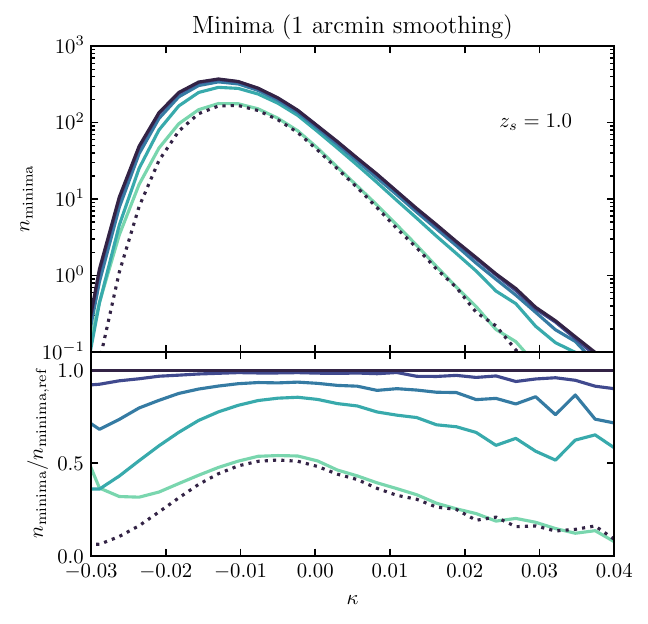}   
    \includegraphics[width=0.325\textwidth]{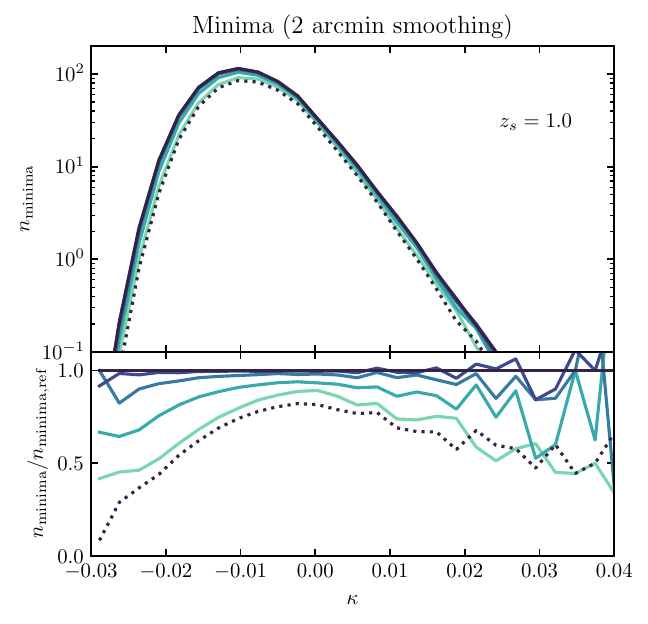}
    \caption{WL convergence PDF (top row), peak counts (center row), and minimum counts (bottom row); computed in the case of 0.5 (left column), 1 (center column), and 2 (right column) arcmin smoothing. All the observables are computed on the B realization of the MTNG740-DM runs taking  $z_s = 1.0$. The solid lines indicate the mean of 1195 $5 \times 5 \, \mathrm{deg}^2$ square maps with increasing darkness representing increasing resolution both in mass and $N_{\mathrm{side}}$. The dotted line refers to the case with the highest mass resolution but down-sampled to $N_{\mathrm{side}}$ = 1024. In each lower sub-panel, we show the ratio with respect to the reference case with $N_{\mathrm{part}} = 4320^3$ and $N_{\mathrm{side}} = 12288$ (noted with the subscript ``ref'').}\label{fig:app_reso}
\end{figure*}

As mentioned in Section~\ref{subsec:observables}, before computing the PDF, peak counts, and minima counts, the convergence maps are smoothed with a Gaussian kernel with a standard deviation of $1 \, \mathrm{arcmin}$. Here we assess how varying the smoothing scale impacts our results from Section~\ref{subsec:resolution}. In Figure~\ref{fig:app_reso} we show the measurements of the PDF (top row), peak counts (center row), and minimum counts (bottom row) for two additional cases, with smoothing scales of 0.5 (left column) and 2 (right column) arcmin. To ease the comparison, we also reproduce again the measurements of these statistics with 1 arcmin smoothing from Figure~\ref{fig:reso} (center column).

We observe that, for all three summary statistics, varying the resolution gives the same qualitative behaviour for the smoothing scales considered here. Quantitatively, we find that when the smoothing scale is smaller, the distortion of the statistics induced by lower angular and mass resolution is greater. The agreement between the dashed and solid green lines is consistent for all smoothing scales, indicating that the conclusions from Section~\ref{subsec:resolution} do not depend on the choice of smoothing scale.

\bsp
\label{lastpage}
\end{document}